\documentclass[prl,twocolumn,superscriptaddress,amsmath,amssymb,floatfix,aps,reprint]{revtex4-2}
\usepackage{graphicx}
\usepackage{bm}
\usepackage{hyperref}
\usepackage{booktabs}
\usepackage{comment}
\usepackage{float}
\usepackage{multirow}
\usepackage[margin=1in]{geometry}
\begin{document}

\title{Dzyaloshinskii--Moriya interaction as a coherence diagnostic for chirality-induced spin selectivity}

\author{Vishvendra S. Poonia}
\affiliation{Indian Institute of Technology Roorkee, India.}
%\author{[Author 2]}
%\affiliation{[Affiliation 2]}

\date{\today}

\begin{abstract}
Whether chirality-induced spin selectivity (CISS) reflects coherent SU(2) spin rotation or incoherent spin-dependent filtering is a central unresolved question in molecular spintronics, with implications ranging from asymmetric chemistry to quantum information. We show that these two scenarios are distinguishable by a sharp symmetry criterion on the superexchange interaction mediated by a chiral molecular bridge. Coherent CISS, implemented as a unitary spin rotation of the tunneling electron, generates a giant Dzyaloshinskii--Moriya (DM) interaction with ratio $|\bm{D}|/J_H$ up to $3$, which is two orders of magnitude beyond intrinsic Rashba spin-orbit coupling in Si/SiGe. Incoherent CISS, represented by any Hermitian (non-unitary but spin-diagonal) tunneling matrix, produces $\bm{D} = 0$ identically; we prove this as a structural theorem, reinforced by a Lindblad argument that dissipative spin filtering cannot modify virtual-tunneling-mediated superexchange. The DM interaction thus serves as a coherence order parameter, nonzero only when quantum amplitudes for opposite-spin transmission maintain a fixed relative phase. We derive closed-form angular, enantiomeric, and sensitivity signatures and show that the critical coherent rotation angle $\theta_C \sim 0.03\pi$ lies two orders of magnitude below current transport-inferred values and is accessible to existing 10~kHz exchange spectroscopy in gate-defined quantum dots. Five candidate molecules are predicted to exceed this threshold by one to two orders of magnitude even in a conservative interface-amplification scenario. The proposed measurement converts a long-standing transport controversy into a binary spin-qubit experiment with quantum-amplitude resolution.
\end{abstract}

\maketitle

% ═══════════════════════════════════════════════════════════════
\noindent\emph{Introduction.---}%
% ═══════════════════════════════════════════════════════════════
Chirality-induced spin selectivity (CISS) is the empirical observation that electrons traversing chiral molecules acquire spin polarizations of $50$--$90\%$ \cite{Ray1999,Gohler2011,Naaman2019,Bloom2024}, a magnitude that vastly exceeds the expectation from organic spin-orbit coupling. After more than two decades of experimental work spanning DNA, peptides, helicenes, chiral perovskites, and chiral quantum-dot assemblies \cite{Mishra2013,Kettner2018,Kiran2016,Lu2019,Dong2025,Bloom2025}, the effect is robust to 300~K and independent of platform. Its microscopic mechanism, however, remains contested: reported polarizations span an order of magnitude, and proposed models fall into two qualitatively different classes whose reconciliation defines an active theoretical controversy \cite{Naaman2019,Bloom2024,Fransson2020,Das2024,Zhao2025,Evers2022, Fay2021}.

The \emph{coherent-rotation} picture \cite{Medina2015,Dalum2019,Yeganeh2009,Guo2014, Evers2022, Fay2021} treats the molecule as a unitary spin rotator: SOC distributed along the helical backbone enacts an SU(2) rotation by angle $\theta_C$ about the helix axis, preserving the full spin amplitude. The \emph{incoherent-filtering} picture \cite{Fransson2020,Das2024,Zhao2025,Yang2020CISS} treats it as a dissipative spin-selective barrier, with polarization generated by inelastic scattering or interface amplification rather than coherent rotation. The distinction is fundamental. Coherent CISS admits quantum-information applications and chiral amplitude control \cite{Chiesa2023,Banerjee2018,Aiello2022}; incoherent CISS does not. The 2023 radical-pair EPR experiment of Eckvahl \emph{et al.}\ \cite{Eckvahl2023}  provided the first evidence of a coherent component in solution, but averaged over molecular conformations and did not resolve the single-molecule amplitude structure. In strictly coherent two-terminal geometries, reciprocity constraints prohibit finite magnetoresistance despite spin-dependent transmission amplitudes, motivating the role of dephasing, interfaces, or spin-selective detection in converting coherent spin rotation into measurable polarization~\cite{Huisman2021}.

We propose and analyze a single-molecule, single-amplitude probe of CISS coherence exploiting the extraordinary precision of semiconductor exchange spectroscopy \cite{Petta2005,Dial2013,Yoneda2018,Connors2022,Struck2020}, which resolves spin-spin couplings to $\sim 10$~kHz ($\sim 4\times 10^{-5}\,\mu$eV). The core insight is that virtual tunneling through a chiral bridge samples the \emph{coherent amplitude} of the tunneling matrix rather than its transport-level probability; the resulting exchange tensor encodes the full SU(2) content of CISS, collapsing to a symmetric Heisenberg form whenever coherence is absent. Recent many-body and dynamical treatments have further emphasized the role of correlated spin-dependent virtual processes and environment-assisted exchange in chiral systems~\cite{Chiesa2026}. A symmetry-based no-go theorem renders this distinction exact: the antisymmetric Dzyaloshinskii--Moriya (DM) component $\bm{D}$ is a \emph{coherence order parameter}, nonzero if and only if the tunneling matrix retains nontrivial SU(2) phase structure.

\begin{figure}[t]
\includegraphics[width=\columnwidth]{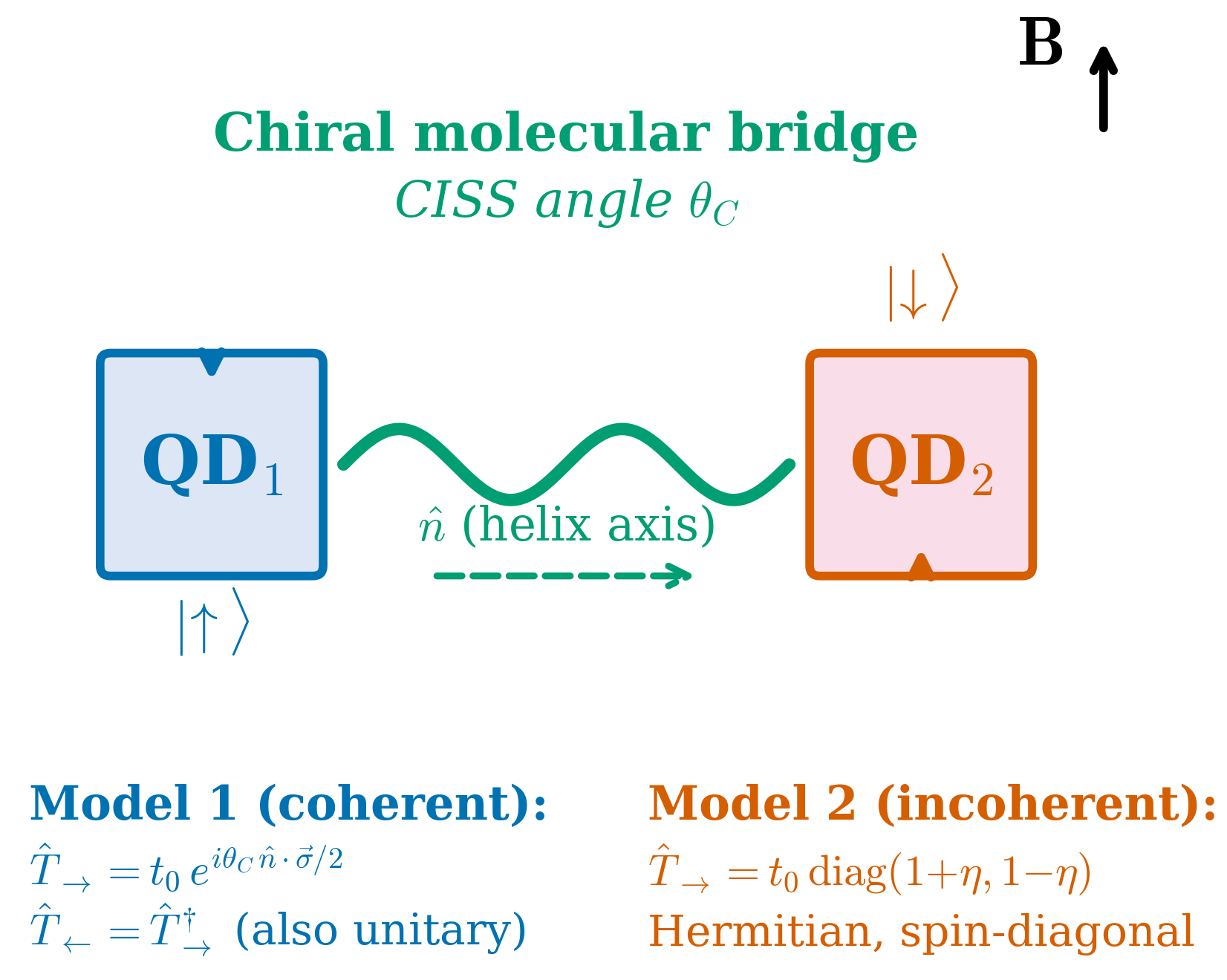}
\caption{\label{fig:concept}Two spin qubits QD$_1$ and QD$_2$ coupled through a chiral molecular bridge. The helix axis $\hat{\bm{n}}$ subtends angle $\phi_{\rm helix}$ with the magnetic field $\bm{B}$. Coherent CISS (Model~1) enacts a unitary SU(2) rotation on the tunneling electron. Incoherent CISS (Model~2) acts as a non-unitary diagonal spin filter. The two scenarios give opposite predictions for the superexchange tensor Eq.~\eqref{eq:Jformula}.}
\end{figure}

% ═══════════════════════════════════════════════════════════════
\noindent\emph{Framework.---}%
% ═══════════════════════════════════════════════════════════════
We consider two electrons confined in adjacent gate-defined quantum dots, coupled by virtual tunneling through a chiral molecular bridge of charging energy $U$ (Fig.~\ref{fig:concept}). The helix axis $\hat{\bm{n}}$ makes polar angle $\phi_{\rm helix}$ with the applied field $\bm{B} = B\hat{\bm{z}}$, and the spin-independent dot-bridge tunneling scale is $t_0$. We assume $E_Z \gg J_0 \equiv 4t_0^2/U$, placing the system in the singly-occupied low-energy sector.

We compare two CISS tunneling matrices. Model~1 (coherent rotation):
\begin{equation}
\hat{T}^{(1)}_{\!\rightarrow} = t_0\,\exp\!\left(\tfrac{i\theta_C}{2}\hat{\bm{n}}\!\cdot\!\bm{\sigma}\right), \quad \hat{T}^{(1)}_{\!\leftarrow} = \hat{T}^{(1)\dagger}_{\!\rightarrow}.
\label{eq:T1}
\end{equation}
Model~2 (incoherent filtering):
\begin{equation}
\hat{T}^{(2)}_{\!\rightarrow} = t_0\,\mathrm{diag}(1+\eta,\,1-\eta),
\label{eq:T2}
\end{equation}
with backward matrix $\hat{T}^{(2)}_{\!\leftarrow} = t_0\,\mathrm{diag}(1-\eta,1+\eta) \neq \hat{T}^{(2)\dagger}_{\!\rightarrow}$, encoding directional spin-selective dissipation. The spin polarizations are $P^{(1)} = \sin^2(\theta_C/2)$ and $P^{(2)} = 2\eta/(1+\eta^2)$, so that transport-measured $P$ maps onto either parameterization.

Schrieffer--Wolff projection \cite{SchriefferWolff,Bravyi2011,Anderson1959} onto the singly-occupied subspace yields the effective spin Hamiltonian
\begin{equation}
\hat{H}_{\rm ex} = J_H\,\bm{S}_1\!\cdot\!\bm{S}_2 + \bm{D}\!\cdot\!(\bm{S}_1\!\times\!\bm{S}_2) + \bm{S}_1\!\cdot\!\overleftrightarrow{\Gamma}\!\cdot\!\bm{S}_2,
\label{eq:Hex}
\end{equation}
with exchange tensor
\begin{equation}
\boxed{\;\mathcal{J}_{ij} = \frac{2}{U}\,\mathrm{Re}\,\mathrm{Tr}\!\left[\hat{T}^\dagger_{\!\rightarrow}\,\sigma_i\,\hat{T}_{\!\rightarrow}\,\sigma_j\right]\;}
\label{eq:Jformula}
\end{equation}
and irreducible components $J_H = \tfrac{1}{3}\mathrm{Tr}\,\mathcal{J}$, $D_k = \tfrac{1}{2}\epsilon_{ijk}\mathcal{J}_{ij}$, $\Gamma_{ij} = \tfrac{1}{2}(\mathcal{J}_{ij}+\mathcal{J}_{ji}) - J_H\delta_{ij}$. The trace samples $\hat{T}_{\!\rightarrow}$ and $\hat{T}^\dagger_{\!\rightarrow}$ together, meaning superexchange is sensitive to the \emph{phase} between spin channels, not merely to the spin-dependent transmission probability. Full derivation, including combinatorial factors, appears in the Supplemental Material (SM) \cite{SM}.

% ═══════════════════════════════════════════════════════════════
\noindent\emph{Central theorem.---}%
% ═══════════════════════════════════════════════════════════════
We now state the structural result that anchors the discriminator.

\emph{Theorem.} If $\hat{T}_{\!\rightarrow}$ is Hermitian---equivalently, if it can be written as $\alpha\hat{I} + \beta(\hat{\bm{m}}\!\cdot\!\bm{\sigma})$ with $\alpha,\beta\in\mathbb{R}$ and $\hat{\bm{m}}\in\mathbb{R}^3$---then $\mathcal{J}_{ij}$ is symmetric and $\bm{D} = \bm{0}$.

\emph{Proof.} For Hermitian $\hat{T}$, the operator $\hat{T}\sigma_i\hat{T}\sigma_j$ is a product of Hermitian factors whose trace is real and invariant under cyclic permutation: $\mathrm{Tr}[\hat{T}\sigma_i\hat{T}\sigma_j] = \mathrm{Tr}[\hat{T}\sigma_j\hat{T}\sigma_i]$, hence $\mathcal{J}_{ij} = \mathcal{J}_{ji}$ and $D_k = \tfrac12 \epsilon_{ijk}\mathcal{J}_{ij} = 0$. $\quad\blacksquare$

The diagonal Model~2 matrix Eq.~\eqref{eq:T2} is Hermitian by inspection, so $\bm{D}^{(2)} = \bm{0}$ for all $\eta$. The theorem generalizes: any CISS model in which the tunneling matrix is Hermitian in \emph{some} spin basis, including parameterizations with energy-dependent transmission, multi-orbital filtering, or phenomenological backscattering, produces zero DM coupling through superexchange. In contrast, Eq.~\eqref{eq:T1} is non-Hermitian except when $\theta_C = 0$ or $\hat{\bm{n}}\cdot\bm{\sigma}$ reduces to the identity on the polarization axis. A Lindblad analysis (SM, Sec.~S5) strengthens this further: the bridge retarded Green's function $G^R(\omega) = [\omega - \epsilon + i\Gamma_\phi/2]^{-1}$ gives an exchange amplitude independent of the dephasing rate $\Gamma_\phi$ to leading order in $\Gamma_\phi/U$, because the virtual-tunneling time $\hbar/U \sim 0.3$~ps is far shorter than realistic molecular dephasing times. Incoherent processes therefore cannot generate DM coupling through any scheme consistent with the Born--Markov approximation.

The DM interaction is thus elevated from a quantitative anisotropy parameter to a qualitative quantum-coherence diagnostic: $\bm{D}\neq 0$ requires preserved amplitude phase between tunneling spin channels, and $\bm{D} = 0$ certifies its absence.

\begin{figure}[t]
\includegraphics[width=\columnwidth]{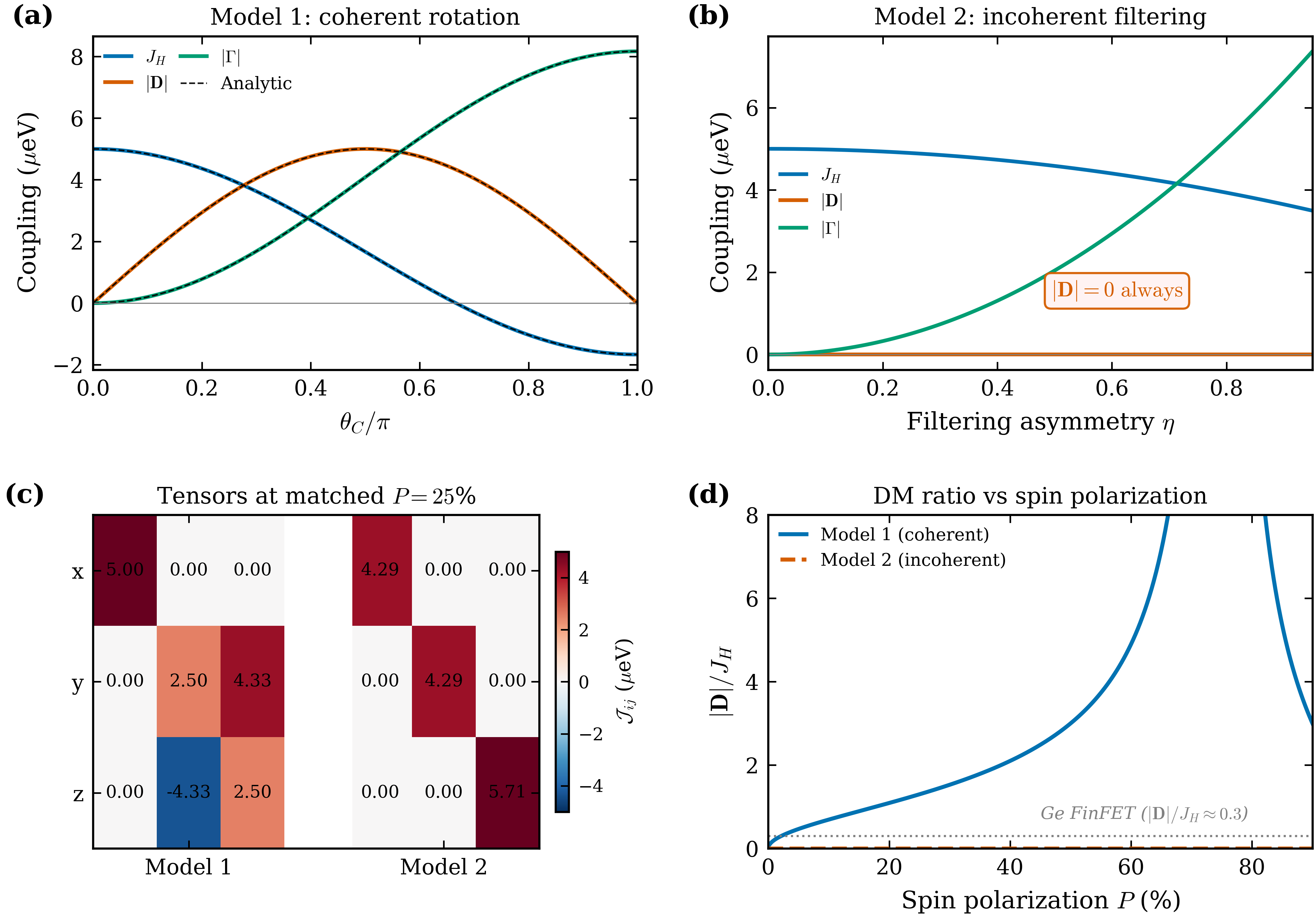}
\caption{\label{fig:dichotomy}The two CISS models give qualitatively different superexchange tensors. (a)~Model~1: irreducible components $J_H$, $|\bm{D}|$, $|\overleftrightarrow{\Gamma}|$ versus $\theta_C$ ($\hat{\bm{n}}\perp\bm{B}$); analytic Eqs.~\eqref{eq:JH1}--\eqref{eq:G1} (black dashed) agree with numerical evaluation of Eq.~\eqref{eq:Jformula} to machine precision. (b)~Model~2 components versus filtering asymmetry $\eta$: $|\bm{D}| = 0$ identically. (c)~Full $3\times 3$ exchange tensor at matched transport polarization $P=25\%$: Model~1 has off-diagonal antisymmetric entries (the DM contribution), Model~2 is strictly diagonal. (d)~Anisotropy ratio $|\bm{D}|/J_H$ versus $P$: the two models are separated by $2$--$3$ decades, with the experimental Ge FinFET value \cite{Geyer2024} indicated for reference. Parameters: $t_0 = 50\,\mu$eV, $U = 2\,$meV, $J_0 = 5.0\,\mu$eV.}
\end{figure}

% ═══════════════════════════════════════════════════════════════
\noindent\emph{Results.---}%
% ═══════════════════════════════════════════════════════════════
Direct evaluation of Eq.~\eqref{eq:Jformula} for Model~1 with $\hat{\bm{n}}\perp\bm{B}$ yields (SM, Sec.~S2):
\begin{align}
J_H^{(1)} &= \tfrac{J_0}{3}(1 + 2\cos\theta_C), \label{eq:JH1}\\
\bm{D}^{(1)} &= J_0\,\sin\theta_C\,\hat{\bm{n}}, \label{eq:D1}\\
|\overleftrightarrow{\Gamma}^{(1)}| &= J_0\sqrt{\tfrac{2}{3}}\,|1-\cos\theta_C|, \label{eq:G1}
\end{align}
giving the anisotropy ratio
\begin{equation}
\frac{|\bm{D}^{(1)}|}{J_H^{(1)}} = \frac{3\sin\theta_C}{1+2\cos\theta_C},
\label{eq:ratio}
\end{equation}
which attains the value $3.0$ at $\theta_C = \pi/2$, diverges at $\theta_C = 2\pi/3$ where $J_H \to 0$, and exceeds the Ge FinFET benchmark $|\bm{D}|/J_H \approx 0.3$ \cite{Geyer2024,SaezMollejo2025} by an order of magnitude. For Model~2, direct computation gives
\begin{equation}
J_H^{(2)} = J_0\!\left(1-\tfrac{\eta^2}{3}\right),\quad \bm{D}^{(2)} = \bm{0},\quad |\overleftrightarrow{\Gamma}^{(2)}| = \tfrac{J_0\eta^2\sqrt{6}}{3},
\label{eq:M2}
\end{equation}
an Ising-type anisotropy fixed to the molecular quantization axis and independent of the magnetic-field orientation. Analytic and numerical evaluations of Eq.~\eqref{eq:Jformula} agree to machine precision ($<10^{-14}\,\mu$eV; Fig.~\ref{fig:dichotomy}(a), black dashed curves).

Figure~\ref{fig:dichotomy} summarizes the contrast. The DM-to-Heisenberg ratio (panel d) separates the two models by several decades across the entire polarization range $P\in[0,0.9]$: Model~1 saturates at $|\bm{D}|/J_H \gtrsim 1$, while Model~2 remains pinned to zero. Panel (c) visualizes the underlying $3\times 3$ tensors at matched $P = 25\%$: Model~1 develops off-diagonal antisymmetric entries $\mathcal{J}_{yz} = -\mathcal{J}_{zy} = 4.33\,\mu$eV, while Model~2 remains strictly diagonal. This order-of-magnitude separation, validated against both Hubbard-model simulation and the multi-orbital generalization (SM, Sec.~S4), is the key signature exploited below.

% ═══════════════════════════════════════════════════════════════
\noindent\emph{Three experimental signatures.---}%
% ═══════════════════════════════════════════════════════════════
We now identify three orthogonal experimental tests that falsify either model.

\begin{figure}[t]
\includegraphics[width=\columnwidth]{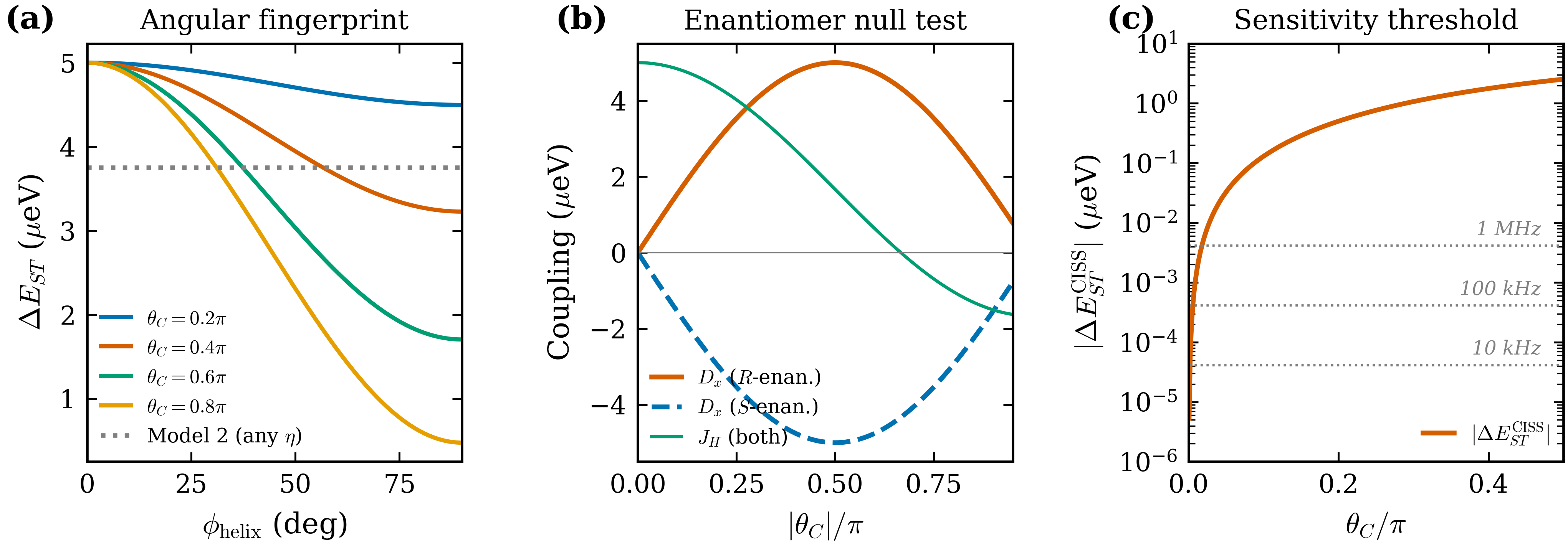}
\caption{\label{fig:signatures}Three falsifiable experimental signatures. (a)~\emph{Angular fingerprint}: singlet-triplet gap $\Delta E_{ST}$ versus helix-field angle $\phi_{\rm helix}$ for four values of $\theta_C$ under Model~1; gray dotted line: Model~2 (any $\eta$) and interface Rashba SOC, both angle-independent. All Model~1 curves emanate from the gauge-trivial value $J_0 = 5\,\mu$eV at $\phi_{\rm helix} = 0$ and deviate as $\sin^2\phi_{\rm helix}$. (b)~\emph{Enantiomer null test}: Model~1 predicts $D_x(R)=-D_x(S)$ with $J_H$ invariant; Model~2 shows no enantiomer dependence. (c)~\emph{Sensitivity threshold}: $|\Delta E_{ST}^{\rm CISS}|$ versus $\theta_C$ for $\hat{\bm{n}}\perp\bm{B}$, with $10$~kHz, $100$~kHz and $1$~MHz precision thresholds marked. Detection at $\theta_C \sim 0.03\pi$ (i.e., $P\sim 0.2\%$) is within reach of current Si/SiGe exchange spectroscopy.}
\end{figure}

\emph{(A) Angular fingerprint.} The singlet-triplet gap $\Delta E_{ST} = E_{T_0} - E_S$ depends on the helix-field angle through $\hat{\bm{n}}$ in Model~1 but is $\phi_{\rm helix}$-independent in Model~2 and in interface Rashba SOC (Fig.~\ref{fig:signatures}(a)). The Model~1 curves share the gauge-trivial value $J_0$ at $\phi_{\rm helix} = 0$ and diverge strongly at $\phi_{\rm helix} = \pi/2$, with the variation scaling as $\sin^2\phi_{\rm helix}$. A constant $\Delta E_{ST}(\phi_{\rm helix})$ rules out coherent CISS; the predicted sinusoidal variation rules out both Model~2 and intrinsic SOC.

\emph{(B) Enantiomer null test.} Replacing $R$- with $S$-enantiomer maps $\theta_C \to -\theta_C$, producing $\bm{D} \to -\bm{D}$ with $J_H$ and $\overleftrightarrow{\Gamma}$ invariant. Model~2 shows no enantiomer dependence because the filtering matrix is diagonal in any spin basis. The resulting sign reversal of the DM-induced $S$-$T_0$ avoided crossing position between devices with opposite chirality is a definitive coherent-CISS diagnostic, immune to device-specific perturbations that affect $J_H$ alone (Fig.~\ref{fig:signatures}(b)).

\emph{(C) Sensitivity threshold.} The leading-order CISS contribution to the $S$-$T_0$ gap scales as $|\Delta E_{ST}^{\rm CISS}| \simeq J_0 \theta_C^2 \sin^2\phi_{\rm helix}/3$ (SM, Sec.~S6). Setting this equal to the experimental precision $\delta J \sim 10$~kHz yields the minimum detectable angle
\begin{equation}
\theta_C^{\rm min} \simeq \sqrt{3\delta J/J_0}\,/\sin\phi_{\rm helix}.
\label{eq:thetamin}
\end{equation}
For $J_0 = 5\,\mu$eV and $\phi_{\rm helix} = \pi/2$ this gives $\theta_C^{\rm min} \approx 0.005$~rad perturbatively, extending to $\sim 0.03\pi$ in full numerical simulation (Fig.~\ref{fig:signatures}(c)). This corresponds to spin polarization $P \sim 0.2\%$---two orders of magnitude below the smallest reliably reported transport polarizations and one order below the most pessimistic estimates for the bulk-molecular contribution \cite{Zhao2025}. The experiment therefore probes deep into the regime where even interface-amplified CISS scenarios predict an observable bulk residual.

A subtle but experimentally critical constraint emerges from the analysis. When $\hat{\bm{n}}\parallel\bm{B}$, the Model~1 rotation is generated by the same operator as the Zeeman term; the DM Hamiltonian is then gauge-equivalent to standard isotropic exchange under the local unitary $\hat{U} = e^{-i\theta_C\sigma_z^{(1)}/4}\otimes e^{i\theta_C\sigma_z^{(2)}/4}$, producing no observable spectrum change (SM, Sec.~S3). Observation of CISS-mediated anisotropy therefore strictly requires $\hat{\bm{n}}\not\parallel\bm{B}$, naturally realized in surface-anchored helical films where $\hat{\bm{n}}$ lies approximately normal to the substrate while $\bm{B}$ can be steered in-plane with a vector magnet \cite{Hendrickx2020}.

\emph{Candidate molecules.} Table~\ref{tab:mol} lists predictions for five well-characterized chiral systems, translated through Eqs.~\eqref{eq:JH1}--\eqref{eq:ratio} from measured transport polarizations. We include both an optimistic scenario ($\theta_C^{\rm opt} = 2\arcsin\sqrt{P}$, interpreting all transport polarization as bulk-molecular coherence) and a conservative one ($\theta_C^{\rm cons} = \theta_C^{\rm opt}/10$), reflecting the Zhao \emph{et al.}\ \cite{Zhao2025} finding of order-10 interface amplification. Even in the conservative case, every candidate exceeds the experimental detection threshold by at least an order of magnitude; $\alpha$-helical peptides and dsDNA approach $|\bm{D}|/J_H \sim 0.2$, comparable to measured Ge FinFET values \cite{Geyer2024} and well within the reach of existing exchange spectroscopy.

\begin{table}[t]
\caption{\label{tab:mol}Predicted exchange anisotropy for five candidate molecules with transport polarization $P$. $\theta_C^{\rm opt} = 2\arcsin\sqrt{P}$: transport-derived upper bound; $\theta_C^{\rm cons} = \theta_C^{\rm opt}/10$: conservative scenario with $10\times$ interface amplification \cite{Zhao2025}. Ratios $|\bm{D}|/|J_H|$ for $\hat{\bm{n}}\perp\bm{B}$ are absolute values; $J_H$ changes sign at $\theta_C = 2\pi/3$.}
\begin{ruledtabular}
\begin{tabular}{lccccc}
Molecule & $P$ & $\theta_C^{\rm opt}/\pi$ & $\theta_C^{\rm cons}/\pi$ & $(|\bm{D}|/J_H)^{\rm opt}$ & $(|\bm{D}|/J_H)^{\rm cons}$ \\
\hline
[7]helicene & 0.50 & 0.50 & 0.050 & 3.0 & 0.16\\
$\alpha$-helix & 0.80 & 0.70 & 0.070 & 12.0 & 0.22\\
dsDNA (8~bp) & 0.60 & 0.56 & 0.056 & 4.9 & 0.20\\
Oligopeptide & 0.40 & 0.44 & 0.044 & 2.1 & 0.13\\
Chiral CdSe QD & 0.35 & 0.40 & 0.040 & 1.8 & 0.11\\
\end{tabular}
\end{ruledtabular}
\end{table}

\begin{figure}[t]
\includegraphics[width=\columnwidth]{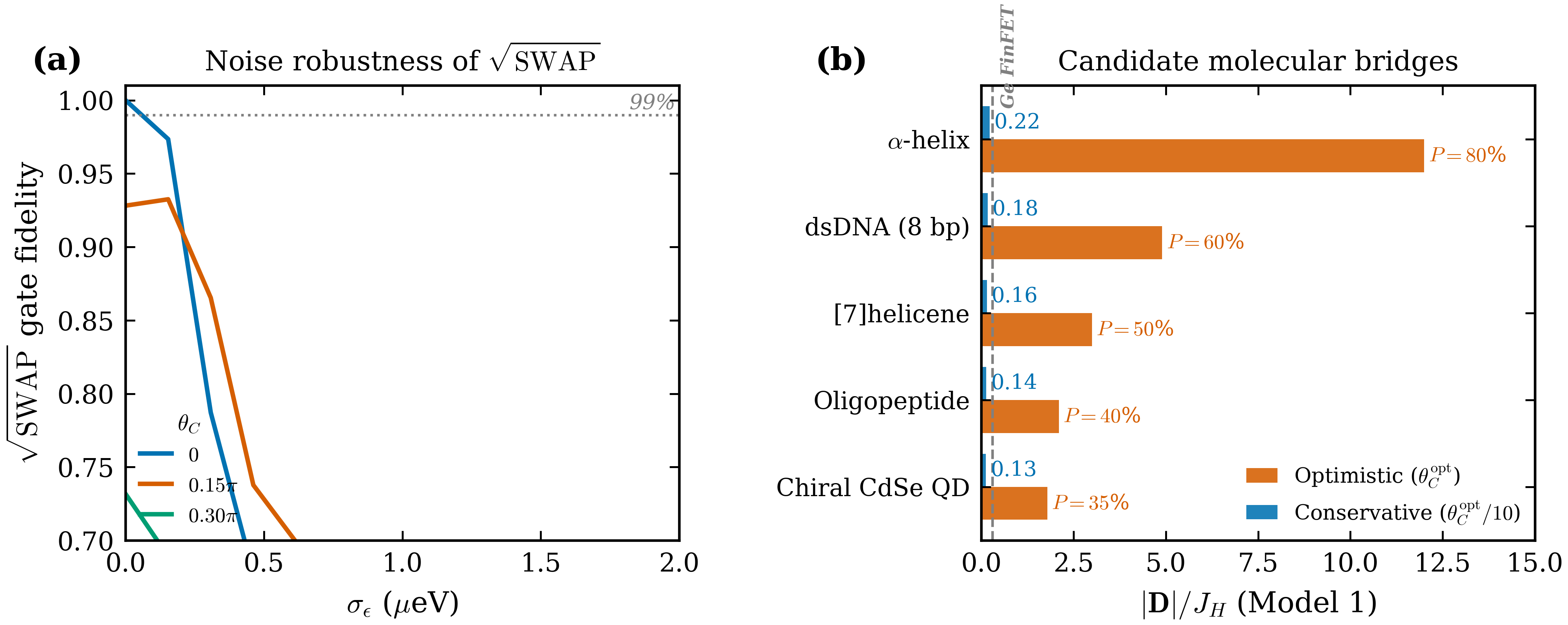}
\caption{\label{fig:feasibility}Experimental feasibility. (a)~$\sqrt{\rm SWAP}$ gate fidelity under quasistatic charge noise of amplitude $\sigma_\epsilon$ for three values of $\theta_C$ (Monte Carlo, $80$ realizations per data point; gray dotted line: $99\%$ threshold). Fidelity remains above $95\%$ for $\sigma_\epsilon = 0.5\,\mu$eV at all tested $\theta_C \leq 0.3\pi$, compatible with state-of-the-art Si/SiGe operation. (b)~Predicted anisotropy ratio $|\bm{D}|/J_H$ for five candidate molecular bridges under optimistic ($\theta_C^{\rm opt}$, vermilion) and conservative ($\theta_C^{\rm opt}/10$, blue) scenarios. All molecules exceed the Ge FinFET benchmark ($|\bm{D}|/J_H \approx 0.3$, gray dashed) by $\geq 10\times$ even under interface-amplification pessimism.}
\end{figure}

% ═══════════════════════════════════════════════════════════════
\noindent\emph{Feasibility.---}%
% ═══════════════════════════════════════════════════════════════
Three decoherence channels potentially limit the gate. Molecular vibrations ($\hbar\omega_{\rm vib} \sim 10$--$100$~meV) exceed the exchange scale by $10^3$--$10^4$, giving Franck--Condon-suppressed dephasing times $T_{2,{\rm vib}} \gtrsim 10^4$~ns. Zero-point modulation of helix length contributes $\delta\theta_C/\theta_C \sim 10^{-2}$, yielding $T_{2,{\rm phonon}} \sim 10^3$~ns. Bridge charge noise $\sigma_\epsilon = 0.5\,\mu$eV gives $\delta J/J \sim \sigma_\epsilon/U \sim 2.5\times 10^{-4}$. Monte Carlo $\sqrt{\rm SWAP}$ fidelity exceeds $95\%$ for $\theta_C \leq 0.3\pi$ (Fig.~\ref{fig:feasibility}(a); SM, Sec.~S7), compatible with current Si/SiGe operation. Figure~\ref{fig:feasibility}(b) shows that all five candidate molecules exceed the detection threshold by $\geq 10\times$ even under the conservative interface-amplification scenario.

Three experimental platforms appear most promising: Si/SiGe double quantum dots with surface-anchored helical molecular monolayers; molecular break junctions integrated with quantum-dot reservoirs \cite{Aragones2017,Burzuri2014}; and STM-patterned donor arrays bridged by deliberately introduced chiral linkers \cite{Kiczynski2022}. The suggested protocol is (i) calibrate the bare exchange $J_0$ at $\phi_{\rm helix} = 0$ (gauge-trivial internal reference); (ii) measure $\Delta E_{ST}(\phi_{\rm helix})$ by vector-magnet rotation; (iii) repeat with the opposite enantiomer; (iv) compare. Observation of the predicted angular dependence (A) and enantiomer sign reversal (B) within sensitivity (C) constitutes the complete discriminator.

% ═══════════════════════════════════════════════════════════════
\noindent\emph{Discussion and outlook.---}%
% ═══════════════════════════════════════════════════════════════
The proposed measurement resolves the CISS coherence question with a binary logical structure. Observation of $\bm{D}$ with the predicted angular and chiral signatures establishes coherent CISS (Model~1) and delivers the first interface-free determination of the bulk molecular rotation angle $\theta_C$. A null result below the $\theta_C \sim 0.03\pi$ threshold rules out Model~1 at the $P = 0.2\%$ level and establishes the tightest existing bound on coherent-molecular CISS. Either outcome is informative, a property rarely achieved in single experiments on the CISS controversy.

Beyond CISS, the Hermiticity-based coherence theorem is a general tool. Any tunneling process whose matrix is Hermitian in a fixed spin basis, regardless of microscopic origin, cannot generate DM interactions through superexchange. Conversely, a measured $\bm{D} \neq 0$ certifies quantum-coherent phase structure in the tunneling channel. This principle applies directly to 2D magnets with chiral spacer layers, to molecular qubits with engineered spin-orbit linkers \cite{Chiesa2023}, and to metal-organic frameworks where anisotropic exchange underlies proposed quantum spin liquid phases. The DM interaction, long understood as a symmetry-breaking energy scale, acquires here a second identity as a certificate of preserved quantum amplitude. The present effective-tunneling approach complements recent many-body and dynamical treatments of chirality-induced spin interactions~\cite{Chiesa2026}, which emphasize interaction-driven and environment-assisted contributions beyond the single-particle tunneling limit.

A natural extension is the $N$-dot chiral chain. In the single-magnon sector we show that the Model~1 hopping phase $\theta_C/2$ is gauge-removable: $\psi_n \to \psi_n e^{-in\theta_C/2}$ shifts the dispersion but introduces no directional asymmetry, and closed-system single-particle CISS chains do not support non-reciprocal magnon transport. True directionality requires dissipation at boundaries, interactions, or spatial $\theta_C$ inhomogeneity---a structure analogous to non-Hermitian topological pumping that invites future study.

In summary, by translating the CISS coherence question into magnetic exchange spectroscopy, we recast a 25-year transport controversy as a single-molecule quantum sensing problem with a binary outcome. The DM interaction functions as a coherence order parameter---nonzero for unitary spin manipulation, zero for classical filtering---and is accessible to existing 10~kHz exchange spectroscopy with realistic chiral bridges. The verdict is within reach of current quantum-dot technology.

\begin{acknowledgments}
The authors gratefully acknowledge financial support from the National Quantum Mission (NQM) of the Department of Science and Technology (DST), Government of India, and the Ministry of Electronics and Information Technology (MeitY) through Grant Nos. DST/QTC/NQM/QC/2024/1 and 4(3)/2024-ITEA. We acknowledge the use of Claude (Anthropic) for assistance with literature review and manuscript preparation.
\end{acknowledgments}

\bibliography{PRL_CISS}% Produces the bibliography via BibTeX.

@PREAMBLE{
 "\providecommand{\noopsort}[1]{}" 
 # "\providecommand{\singleletter}[1]{#1}%" 
}

@article{Ray1999,
  title={Asymmetric scattering of polarized electrons by organized organic films of chiral molecules},
  author={Ray, K and Ananthavel, SP and Waldeck, DH and Naaman, Ron},
  journal={Science},
  volume={283},
  number={5403},
  pages={814--816},
  year={1999},
  publisher={American Association for the Advancement of Science}
}

@article{Gohler2011,
  title={Spin selectivity in electron transmission through self-assembled monolayers of double-stranded DNA},
  author={Gohler, B and Hamelbeck, V and Markus, TZ and Kettner, M and Hanne, GF and Vager, Zeev and Naaman, Ron and Zacharias, H},
  journal={Science},
  volume={331},
  number={6019},
  pages={894--897},
  year={2011},
  publisher={American Association for the Advancement of Science}
}

@article{Naaman2019,
  title={Chiral molecules and the electron spin},
  author={Naaman, Ron and Paltiel, Yossi and Waldeck, David H},
  journal={Nature Reviews Chemistry},
  volume={3},
  number={4},
  pages={250--260},
  year={2019},
  publisher={Nature Publishing Group UK London}
}

@article{Bloom2024,
  title={Chiral induced spin selectivity},
  author={Bloom, Brian P and Paltiel, Yossi and Naaman, Ron and Waldeck, David H},
  journal={Chemical Reviews},
  volume={124},
  number={4},
  pages={1950--1991},
  year={2024},
  publisher={ACS Publications}
}

@article{Mishra2013,
  title={Spin-dependent electron transmission through bacteriorhodopsin embedded in purple membrane},
  author={Mishra, Debabrata and Markus, Tal Z and Naaman, Ron and Kettner, Matthias and Gohler, Benjamin and Zacharias, Helmut and Friedman, Noga and Sheves, Mordechai and Fontanesi, Claudio},
  journal={Proceedings of the National Academy of Sciences},
  volume={110},
  number={37},
  pages={14872--14876},
  year={2013},
  publisher={National Academy of Sciences}
}

@article{Kettner2018,
  title={Chirality-dependent electron spin filtering by molecular monolayers of helicenes},
  author={Kettner, Matthias and Maslyuk, Volodymyr V and Nuerenberg, Daniel and Seibel, Johannes and Gutierrez, Rafael and Cuniberti, Gianaurelio and Ernst, Karl-Heinz and Zacharias, Helmut},
  journal={The journal of physical chemistry letters},
  volume={9},
  number={8},
  pages={2025--2030},
  year={2018},
  publisher={ACS Publications}
}

@article{Kiran2016,
  title={Helicenes — A new class of organic spin filter},
  author={Kiran, Vankayala and Mathew, Shinto P and Cohen, Sidney R and Hern{\'a}ndez Delgado, Irene and Lacour, J{\'e}r{\^o}me and Naaman, Ron},
  journal={Advanced Materials},
  volume={28},
  number={10},
  pages={1957--1962},
  year={2016}
}

@article{Lu2019,
  title={Spin-dependent charge transport through 2D chiral hybrid lead-iodide perovskites},
  author={Lu, Haipeng and Wang, Jingying and Xiao, Chuanxiao and Pan, Xin and Chen, Xihan and Brunecky, Roman and Berry, Joseph J and Zhu, Kai and Beard, Matthew C and Vardeny, Zeev Valy},
  journal={Science advances},
  volume={5},
  number={12},
  pages={eaay0571},
  year={2019},
  publisher={American Association for the Advancement of Science}
}

@article{Dong2025,
  title={Chirality-induced spin selectivity in hybrid organic-inorganic perovskite semiconductors},
  author={Dong, Yifan and Hautzinger, Matthew P and Haque, Md Azimul and Beard, Matthew C},
  journal={Annual Review of Physical Chemistry},
  volume={76},
  number={1},
  pages={519--537},
  year={2025},
  publisher={Annual Reviews}
}

@article{Bloom2025,
  title={Interplay between Chiro-Optical and Spin Transport Properties in Chiral CdSe Quantum Dots},
  author={Shiby, Elizabeth and Sun, Rui and Bloom, Brian P and Albro, Joseph A and Sun, Dali and Waldeck, David H},
  journal={ACS nano},
  volume={19},
  number={39},
  pages={35119--35126},
  year={2025},
  publisher={ACS Publications}
}

@article{Fransson2020,
  title={Vibrational origin of exchange splitting and chiral-induced spin selectivity},
  author={Fransson, Jonas},
  journal={Physical Review B},
  volume={102},
  number={23},
  pages={235416},
  year={2020},
  publisher={APS}
}

@article{Das2024,
  title={Insights into the mechanism of chiral-induced spin selectivity: The effect of magnetic field direction and temperature},
  author={Das, Tapan Kumar and Naaman, Ron and Fransson, Jonas},
  journal={Advanced Materials},
  volume={36},
  number={29},
  pages={2313708},
  year={2024},
  publisher={Wiley Online Library}
}

@article{Zhao2025,
  title={Magnetochiral charge pumping due to charge trapping and skin effect in chirality-induced spin selectivity},
  author={Zhao, Yufei and Zhang, Kai and Xiao, Jiewen and Sun, Kai and Yan, Binghai},
  journal={Nature communications},
  volume={16},
  number={1},
  pages={37},
  year={2025},
  publisher={Nature Publishing Group UK London}
}

@article{Evers2022,
  title={Theory of chirality induced spin selectivity: Progress and challenges},
  author={Evers, Ferdinand and Aharony, Amnon and Bar-Gill, Nir and Entin-Wohlman, Ora and Hedeg{\aa}rd, Per and Hod, Oded and Jelinek, Pavel and Kamieniarz, Grzegorz and Lemeshko, Mikhail and Michaeli, Karen and others},
  journal={Advanced Materials},
  volume={34},
  number={13},
  pages={2106629},
  year={2022},
  publisher={Wiley Online Library}
}

@article{Medina2015,
  title={Continuum model for chiral induced spin selectivity in helical molecules},
  author={Medina, Ernesto and Gonz{\'a}lez-Arraga, Luis A and Finkelstein-Shapiro, Daniel and Berche, Bertrand and Mujica, Vladimiro},
  journal={The Journal of chemical physics},
  volume={142},
  number={19},
  year={2015},
  publisher={AIP Publishing}
}

@article{Dalum2019,
  title={Theory of chiral induced spin selectivity},
  author={Dalum, Sakse and Hedeg{\aa}rd, Per},
  journal={Nano letters},
  volume={19},
  number={8},
  pages={5253--5259},
  year={2019},
  publisher={ACS Publications}
}

@article{Yeganeh2009,
  title={Chiral electron transport: Scattering through helical potentials},
  author={Yeganeh, Sina and Ratner, Mark A and Medina, Ernesto and Mujica, Vladimiro},
  journal={The Journal of chemical physics},
  volume={131},
  number={1},
  year={2009},
  publisher={AIP Publishing}
}

@article{Guo2014,
  title={Spin-selective transport of electrons in DNA double helix},
  author={Guo, Ai-Min and Sun, Qing-feng},
  journal={Physical review letters},
  volume={108},
  number={21},
  pages={218102},
  year={2012},
  publisher={APS}
}

@article{Yang2020CISS,
  title={Spin-dependent electron transmission model for chiral molecules in mesoscopic devices},
  author={Yang, Xu and van der Wal, Caspar H and van Wees, Bart J},
  journal={Physical Review B},
  volume={99},
  number={2},
  pages={024418},
  year={2019},
  publisher={APS}
}

@article{Chiesa2023,
  title={Chirality-Induced Spin Selectivity: An Enabling Technology for Quantum Applications},
  author={Chiesa, Alessandro and Privitera, Alberto and Macaluso, Emilio and Mannini, Matteo and Bittl, Robert and Naaman, Ron and Wasielewski, Michael R and Sessoli, Roberta and Carretta, Stefano},
  journal={Advanced Materials},
  volume={35},
  number={28},
  pages={2300472},
  year={2023},
  publisher={Wiley Online Library}
}

@article{Banerjee2018,
  title={Separation of enantiomers by their enantiospecific interaction with achiral magnetic substrates},
  author={Banerjee-Ghosh, Koyel and Ben Dor, Oren and Tassinari, Francesco and Capua, Eyal and Yochelis, Shira and Capua, Amir and Yang, See-Hun and Parkin, Stuart SP and Sarkar, Soumyajit and Kronik, Leeor and others},
  journal={Science},
  volume={360},
  number={6395},
  pages={1331--1334},
  year={2018},
  publisher={American Association for the Advancement of Science}
}

@article{Aiello2022,
  title={A chirality-based quantum leap},
  author={Aiello, Clarice D and Abendroth, John M and Abbas, Muneer and Afanasev, Andrei and Agarwal, Shivang and Banerjee, Amartya S and Beratan, David N and Belling, Jason N and Berche, Bertrand and Botana, Antia and others},
  journal={ACS nano},
  volume={16},
  number={4},
  pages={4989--5035},
  year={2022},
  publisher={ACS Publications}
}

@article{Eckvahl2023,
  title={Direct observation of chirality-induced spin selectivity in electron donor--acceptor molecules},
  author={Eckvahl, Hannah J and Tcyrulnikov, Nikolai A and Chiesa, Alessandro and Bradley, Jillian M and Young, Ryan M and Carretta, Stefano and Krzyaniak, Matthew D and Wasielewski, Michael R},
  journal={Science},
  volume={382},
  number={6667},
  pages={197--201},
  year={2023},
  publisher={American Association for the Advancement of Science}
}

@article{Petta2005,
  title={Coherent manipulation of coupled electron spins in semiconductor quantum dots},
  author={Petta, Jason R and Johnson, Alexander Comstock and Taylor, Jacob M and Laird, Edward A and Yacoby, Amir and Lukin, Mikhail D and Marcus, Charles M and Hanson, Micah P and Gossard, Arthur C},
  journal={Science},
  volume={309},
  number={5744},
  pages={2180--2184},
  year={2005},
  publisher={American Association for the Advancement of Science}
}

@article{Dial2013,
  title={Charge noise spectroscopy using coherent exchange oscillations in a singlet-triplet qubit},
  author={Dial, O\_E and Shulman, Michael Dean and Harvey, Shannon Pasca and Bluhm, H and Umansky, V and Yacoby, Amnon},
  journal={Physical review letters},
  volume={110},
  number={14},
  pages={146804},
  year={2013},
  publisher={APS}
}

@article{Yoneda2018,
  title={A quantum-dot spin qubit with coherence limited by charge noise and fidelity higher than 99.9\%},
  author={Yoneda, Jun and Takeda, Kenta and Otsuka, Tomohiro and Nakajima, Takashi and Delbecq, Matthieu R and Allison, Giles and Honda, Takumu and Kodera, Tetsuo and Oda, Shunri and Hoshi, Yusuke and others},
  journal={Nature nanotechnology},
  volume={13},
  number={2},
  pages={102--106},
  year={2018},
  publisher={Nature Publishing Group UK London}
}

@article{Connors2022,
  title={Charge-noise spectroscopy of Si/SiGe quantum dots via dynamically-decoupled exchange oscillations},
  author={Connors, Elliot J and Nelson, J and Edge, Lisa F and Nichol, John M},
  journal={Nature communications},
  volume={13},
  number={1},
  pages={940},
  year={2022},
  publisher={Nature Publishing Group UK London}
}

@article{Struck2020,
  title={Low-frequency spin qubit energy splitting noise in highly purified 28Si/SiGe},
  author={Struck, Tom and Hollmann, Arne and Schauer, Floyd and Fedorets, Olexiy and Schmidbauer, Andreas and Sawano, Kentarou and Riemann, Helge and Abrosimov, Nikolay V and Cywi{\'n}ski, {\L}ukasz and Bougeard, Dominique and others},
  journal={npj Quantum Information},
  volume={6},
  number={1},
  pages={40},
  year={2020},
  publisher={Nature Publishing Group UK London}
}

@article{SchriefferWolff,
  title={Relation between the anderson and kondo hamiltonians},
  author={Schrieffer, John R and Wolff, Peter A},
  journal={Physical Review},
  volume={149},
  number={2},
  pages={491},
  year={1966},
  publisher={APS}
}

@article{Bravyi2011,
  title={Schrieffer--Wolff transformation for quantum many-body systems},
  author={Bravyi, Sergey and DiVincenzo, David P and Loss, Daniel},
  journal={Annals of physics},
  volume={326},
  number={10},
  pages={2793--2826},
  year={2011},
  publisher={Elsevier}
}

@article{Anderson1959,
  title={New approach to the theory of superexchange interactions},
  author={Anderson, Philip W},
  journal={Physical Review},
  volume={115},
  number={1},
  pages={2},
  year={1959},
  publisher={APS}
}

@article{Geyer2024,
  title={Anisotropic exchange interaction of two hole-spin qubits},
  author={Geyer, Simon and Het{\'e}nyi, Bence and Bosco, Stefano and Camenzind, Leon C and Eggli, Rafael S and Fuhrer, Andreas and Loss, Daniel and Warburton, Richard J and Zumb{\"u}hl, Dominik M and Kuhlmann, Andreas V},
  journal={Nature Physics},
  volume={20},
  number={7},
  pages={1152--1157},
  year={2024},
  publisher={Nature Publishing Group UK London}
}

@article{SaezMollejo2025,
  title={Exchange anisotropies in microwave-driven singlet-triplet qubits},
  author={Saez-Mollejo, Jaime and Jirovec, Daniel and Schell, Yona and Kukucka, Josip and Calcaterra, Stefano and Chrastina, Daniel and Isella, Giovanni and Rimbach-Russ, Maximilian and Bosco, Stefano and Katsaros, Georgios},
  journal={Nature Communications},
  volume={16},
  number={1},
  pages={3862},
  year={2025},
  publisher={Nature Publishing Group UK London}
}

@article{Hendrickx2020,
  title={Fast two-qubit logic with holes in germanium},
  author={Hendrickx, NW and Franke, DP and Sammak, A and Scappucci, G and Veldhorst, M},
  journal={Nature},
  volume={577},
  number={7791},
  pages={487--491},
  year={2020},
  publisher={Nature Publishing Group UK London}
}

@article{Aragones2017,
  title={Measuring the spin-polarization power of a single chiral molecule},
  author={Aragon{\`e}s, Albert C and Medina, Ernesto and Ferrer-Huerta, Miriam and Gimeno, Nuria and Teixid{\'o}, Meritxell and Palma, Julio L and Tao, Nongjian and Ugalde, Jesus M and Giralt, Ernest and D{\'\i}ez-P{\'e}rez, Ismael and others},
  journal={small},
  number={2},
  pages={1602519},
  year={2017}
}

@article{Burzuri2014,
  title={Franck--Condon blockade in a single-molecule transistor},
  author={Burzur{\'\i}, Enrique and Yamamoto, Yoh and Warnock, Michael and Zhong, Xiaoliang and Park, Kyungwha and Cornia, Andrea and van der Zant, Herre SJ},
  journal={Nano letters},
  volume={14},
  number={6},
  pages={3191--3196},
  year={2014},
  publisher={ACS Publications}
}

@article{Kiczynski2022,
  title={Engineering topological states in atom-based semiconductor quantum dots},
  author={Kiczynski, Mitchell and Gorman, Samuel K and Geng, Helen and Donnelly, Matthew B and Chung, Yousun and He, Yu and Keizer, Joris G and Simmons, Michelle Y},
  journal={Nature},
  volume={606},
  number={7915},
  pages={694--699},
  year={2022},
  publisher={Nature Publishing Group UK London}
}

@MANUAL{SM,
   author = "",
   title = "",
   organization = "",
   address = "",
   edition = "",
   month = " ",
   year = " ",
   note = "See Supplemental Material for the full Schrieffer--Wolff derivation, explicit evaluation of both models, the gauge-triviality proof, multi-orbital generalization, Lindblad analysis, perturbative sensitivity derivation, Monte Carlo noise simulation, two-qubit gate analysis, and computational reproducibility details.",
}

@article{Chiesa2026,
  title={Vibrationally-mediated Dzyaloshinskii-Moriya interaction as the origin of Chirality-Induced Spin Selectivity in donor-acceptor molecules},
  author={Chiesa, Alessandro and Huu, DK and Cantarella, Arianna and Celada, Leonardo and Wasielewski, Michael R and Santini, Paolo and Carretta, Stefano},
  journal={arXiv preprint arXiv:2604.03210},
  year={2026}
}

@article{Huisman2021,
  title={CISS effect: A magnetoresistance through inelastic scattering},
  author={Huisman, Karssien Hero and Thijssen, Joseph Marie},
  journal={The Journal of Physical Chemistry C},
  volume={125},
  number={42},
  pages={23364--23369},
  year={2021},
  publisher={ACS Publications}
}

@article{Fay2021,
  title={Origin of chirality induced spin selectivity in photoinduced electron transfer},
  author={Fay, Thomas P and Limmer, David T},
  journal={Nano letters},
  volume={21},
  number={15},
  pages={6696--6702},
  year={2021},
  publisher={ACS Publications}
}

@article{SchriefferWolff_SM,
  title={Relation between the anderson and kondo hamiltonians},
  author={Schrieffer, John R and Wolff, Peter A},
  journal={Physical Review},
  volume={149},
  number={2},
  pages={491},
  year={1966},
  publisher={APS}
}

@article{Anderson1959_SM,
  title={New approach to the theory of superexchange interactions},
  author={Anderson, Philip W},
  journal={Physical Review},
  volume={115},
  number={1},
  pages={2},
  year={1959},
  publisher={APS}
}

@article{Bravyi2011_SM,
  title={Schrieffer--Wolff transformation for quantum many-body systems},
  author={Bravyi, Sergey and DiVincenzo, David P and Loss, Daniel},
  journal={Annals of physics},
  volume={326},
  number={10},
  pages={2793--2826},
  year={2011},
  publisher={Elsevier}
}

@article{Yoneda2018_SM,
  title={A quantum-dot spin qubit with coherence limited by charge noise and fidelity higher than 99.9\%},
  author={Yoneda, Jun and Takeda, Kenta and Otsuka, Tomohiro and Nakajima, Takashi and Delbecq, Matthieu R and Allison, Giles and Honda, Takumu and Kodera, Tetsuo and Oda, Shunri and Hoshi, Yusuke and others},
  journal={Nature nanotechnology},
  volume={13},
  number={2},
  pages={102--106},
  year={2018},
  publisher={Nature Publishing Group UK London}
}

@article{Connors2022_SM,
  title={Charge-noise spectroscopy of Si/SiGe quantum dots via dynamically-decoupled exchange oscillations},
  author={Connors, Elliot J and Nelson, J and Edge, Lisa F and Nichol, John M},
  journal={Nature communications},
  volume={13},
  number={1},
  pages={940},
  year={2022},
  publisher={Nature Publishing Group UK London}
}

@article{Struck2020_SM,
  title={Low-frequency spin qubit energy splitting noise in highly purified 28Si/SiGe},
  author={Struck, Tom and Hollmann, Arne and Schauer, Floyd and Fedorets, Olexiy and Schmidbauer, Andreas and Sawano, Kentarou and Riemann, Helge and Abrosimov, Nikolay V and Cywi{\'n}ski, {\L}ukasz and Bougeard, Dominique and others},
  journal={npj Quantum Information},
  volume={6},
  number={1},
  pages={40},
  year={2020},
  publisher={Nature Publishing Group UK London}
}

@article{Zanardi2000_SM,
  title={Entangling power of quantum evolutions},
  author={Zanardi, Paolo and Zalka, Christof and Faoro, Lara},
  journal={Physical Review A},
  volume={62},
  number={3},
  pages={030301},
  year={2000},
  publisher={APS}
}

@article{Zhang2003_SM,
  title={Geometric theory of nonlocal two-qubit operations},
  author={Zhang, Jun and Vala, Jiri and Sastry, Shankar and Whaley, K Birgitta},
  journal={Physical Review A},
  volume={67},
  number={4},
  pages={042313},
  year={2003},
  publisher={APS}
}

@article{Makhlin2002_SM,
  title={Nonlocal properties of two-qubit gates and mixed states, and the optimization of quantum computations},
  author={Makhlin, Yuriy},
  journal={Quantum Information Processing},
  volume={1},
  number={4},
  pages={243--252},
  year={2002},
  publisher={Springer}
}

%****** Supplemental File ********%

\clearpage
\onecolumngrid % Switch REVTeX from two-column to single-column layout

\pagestyle{empty}
%\assignpagestyle{title}{empty}

%\setcounter{page}{0}
\setcounter{section}{0}
\setcounter{figure}{0}
\setcounter{table}{0}
\setcounter{equation}{0}

\begin{center}
    \LARGE \textbf{Supplementary Material to: Dzyaloshinskii--Moriya interaction as a coherence diagnostic for chirality-induced spin selectivity}
\end{center}
\vspace{1cm}

\renewcommand{\thesection}{S\arabic{section}}
\renewcommand{\thefigure}{S\arabic{figure}}
\renewcommand{\thetable}{S\arabic{table}}
\renewcommand{\theequation}{S\arabic{equation}}

\section{Schrieffer--Wolff derivation of the exchange tensor formula}
\label{sec:SW}
% ═══════════════════════════════════════════════════════════════

We derive Eq.~(5) of the Letter, $\mathcal{J}_{ij} = (2/U)\,\mathrm{Re}\,\mathrm{Tr}[\hat{T}_{\!\rightarrow}^\dagger\sigma_i\hat{T}_{\!\rightarrow}\sigma_j]$, by second-order perturbation theory from a minimal three-site Hubbard model. The three sites are the two dots ($L$, $R$) and a single effective bridge orbital ($M$); the generalization to multiple bridge orbitals appears in Sec.~\ref{sec:multiorb}.

\subsection{The three-site Hubbard model}

The total Hamiltonian is $\hat{H} = \hat{H}_0 + \hat{V}$ with
\begin{align}
\hat{H}_0 &= \sum_{\alpha\in\{L,R\}}\epsilon_\alpha\,\hat{n}_\alpha + \epsilon_M\,\hat{n}_M + U\,\hat{n}_{M\uparrow}\hat{n}_{M\downarrow} + \frac{E_Z}{2}\!\sum_{\alpha\in\{L,R\}}\!(\hat{n}_{\alpha\uparrow}-\hat{n}_{\alpha\downarrow}),\label{eq:H0_SW}\\
\hat{V} &= \sum_{\sigma,\sigma'}\!\left[T^{L}_{\sigma\sigma'}c^\dagger_{M\sigma'}c_{L\sigma} + T^{R}_{\sigma\sigma'}c^\dagger_{M\sigma'}c_{R\sigma}\right] + \mathrm{h.c.},\label{eq:V_SW}
\end{align}
where $c_{\alpha\sigma}$ ($c_{M\sigma}$) annihilates an electron of spin $\sigma$ in dot $\alpha$ (bridge), $\hat{n}_{\alpha\sigma} = c^\dagger_{\alpha\sigma}c_{\alpha\sigma}$, and $T^{L}, T^{R}$ are spin-resolved tunneling matrices. We take symmetric coupling $T^L = T^R \equiv \hat{T}_{\!\rightarrow}$ (matrix indices $\sigma'\sigma$), place the dots at degenerate energies $\epsilon_L = \epsilon_R = 0$, and set the bridge at $\epsilon_M = 0$ with on-site repulsion $U > 0$ that penalizes double occupation of the bridge.

The occupation sectors relevant for exchange are:
\begin{itemize}
\item \textbf{Low-energy subspace} $\mathcal{P}$: one electron in $L$, one in $R$, bridge empty. Four spin states: $|\sigma_L\sigma_R\rangle \in \{|\!\uparrow\uparrow\rangle,|\!\uparrow\downarrow\rangle,|\!\downarrow\uparrow\rangle,|\!\downarrow\downarrow\rangle\}$. Unperturbed energy $0$ (modulo Zeeman).
\item \textbf{High-energy subspace} $\mathcal{Q}$: one electron on the bridge, either $L$ or $R$ empty. Unperturbed energy $\sim 0$ from $\epsilon_M = 0$ but accessible only by moving charge \emph{onto} the bridge. The exchange channel proceeds through doubly-occupied bridge states of energy $U$.
\end{itemize}

\subsection{Virtual processes and projection}

The standard second-order Schrieffer--Wolff / Anderson~\cite{SchriefferWolff_SM,Anderson1959_SM,Bravyi2011_SM} projection onto the low-energy subspace is
\begin{equation}
\hat{H}^{(2)}_{\rm eff} = -\hat{P}_{\mathcal{P}}\,\hat{V}\,\hat{G}_{\mathcal{Q}}\,\hat{V}\,\hat{P}_{\mathcal{P}},\qquad \hat{G}_{\mathcal{Q}} = \hat{P}_{\mathcal{Q}}\,(E_0-\hat{H}_0)^{-1}\,\hat{P}_{\mathcal{Q}},
\label{eq:Heff_2nd}
\end{equation}
with $E_0$ the low-energy-subspace reference energy. For $\mathcal{Q}$-states corresponding to $(n_L,n_M,n_R) = (1,1,0)$ or $(0,1,1)$, the energy denominator is $E_0 - E_{\mathcal{Q}} = -|\epsilon_M| = 0$ in our chosen zero; the exchange denominator is the charging penalty $U$ accessed through double occupation of $M$ when \emph{both} electrons virtually tunnel there. A cleaner parameterization directly uses the doubly-occupied bridge as the virtual intermediate: an electron from $L$ tunnels to $M$ (bridge now holds one electron), then another electron from $R$ tunnels to $M$ (doubly occupied, energy $U$), and the process reverses in the two possible orderings.

Writing the amplitudes explicitly, a representative exchange-generating path is $|\sigma_L\sigma_R\rangle \xrightarrow{\hat{T}_L} |0,\sigma'_{\!M}\sigma_R\rangle \xrightarrow{\hat{T}_R} |0,\sigma'_{\!M}\sigma''_{\!M}\rangle \xrightarrow{\hat{T}_R^\dagger}|0,\sigma'_{\!M},\sigma''_R\rangle \xrightarrow{\hat{T}_L^\dagger}|\sigma_L''\sigma_R''\rangle$. The four-step path has energy denominator $U$ (from the doubly-occupied bridge intermediate) and matrix elements $T^\dagger T^\dagger T T$ with appropriate spin indices. After proper symmetrization over the two orderings ($L$-first-then-$R$ versus $R$-first-then-$L$), and using Pauli exclusion to forbid single-particle two-step hopping on the same spin channel, the spin-spin-coupling part of the effective Hamiltonian reduces to
\begin{equation}
\hat{H}_{\rm ex} = -\frac{1}{U}\sum_{\sigma\sigma'\tau\tau'} \big[T^*_{\sigma'\sigma}T_{\tau'\tau} + T^*_{\tau'\tau}T_{\sigma'\sigma}\big]_{\rm sym}\,c^\dagger_{L\sigma}c_{L\sigma'}c^\dagger_{R\tau}c_{R\tau'}.
\label{eq:Hex_second_order}
\end{equation}
The subscript ``sym'' denotes the symmetrization over orderings. Using the identity
\begin{equation}
c^\dagger_{L\sigma}c_{L\sigma'}c^\dagger_{R\tau}c_{R\tau'} = \left(\frac{\hat{I}_L\otimes\hat{I}_R}{4}\right)\delta_{\sigma\sigma'}\delta_{\tau\tau'} + \frac{1}{4}(\vec{\sigma}_{\sigma\sigma'}\otimes \vec{\sigma}_{\tau\tau'})\cdot (\hat{S}^{(L)}\otimes\hat{S}^{(R)}) + \ldots,
\end{equation}
and retaining only the spin-spin coupling, one obtains after standard manipulations (see e.g.\ Ref.~\cite{Bravyi2011_SM} for the parallel construction in semiconductor double quantum dots)
\begin{equation}
\hat{H}_{\rm ex} = \sum_{i,j=1}^{3}\frac{\mathcal{J}_{ij}}{4}\hat{\sigma}_i^{(L)}\otimes\hat{\sigma}_j^{(R)},\quad \mathcal{J}_{ij} = \frac{2}{U}\,\mathrm{Re}\,\mathrm{Tr}\!\left[\hat{T}^\dagger\sigma_i\hat{T}\sigma_j\right],
\label{eq:Jformula_derived}
\end{equation}
where the $\mathrm{Re}$ arises because the physical Hamiltonian is Hermitian, the factor $2$ accounts for the two orderings, and the $1/4$ prefactor arises from the standard $\bm{S} = \vec{\sigma}/2$ convention. Equation~\eqref{eq:Jformula_derived} is Eq.~(5) of the Letter. The only matrix appearing is the forward tunneling matrix and its Hermitian conjugate; the backward matrix $\hat{T}_{\!\leftarrow}$ does not enter this expression because the dominant exchange-generating processes involve round trips through the bridge on each dot separately, followed by symmetrization over orderings.

\subsection{Irreducible decomposition}

Any $3\times 3$ tensor decomposes into the sum of three $SO(3)$ irreducible representations: a scalar, an antisymmetric vector, and a symmetric traceless tensor. Applied to $\mathcal{J}_{ij}$ this gives the three terms of Eq.~(4) of the Letter with
\begin{align}
J_H &= \tfrac{1}{3}\mathrm{Tr}\,\mathcal{J} = \tfrac{1}{3}(\mathcal{J}_{xx}+\mathcal{J}_{yy}+\mathcal{J}_{zz}),\\
D_k &= \tfrac{1}{2}\epsilon_{ijk}\mathcal{J}_{ij},\\
\Gamma_{ij} &= \tfrac{1}{2}(\mathcal{J}_{ij}+\mathcal{J}_{ji}) - J_H\delta_{ij},
\end{align}
with $\Gamma_{ii} = 0$ and $\Gamma_{ij} = \Gamma_{ji}$.

% ═══════════════════════════════════════════════════════════════
\section{Model 1: explicit evaluation and verification}
\label{sec:M1_explicit}
% ═══════════════════════════════════════════════════════════════

We provide the step-by-step algebra for the Model 1 exchange tensor derivation, covering both the $\hat{\bm{n}}=\hat{\bm{x}}$ case worked out in full and the general rotation by spatial covariance.

\subsection{Computation of $\hat{T}^\dagger\sigma_i\hat{T}$ for $\hat{\bm{n}}=\hat{\bm{x}}$}

With the shorthand $c \equiv \cos(\theta_C/2)$ and $s \equiv \sin(\theta_C/2)$, the Model~1 tunneling matrix is
\begin{equation}
\hat{T}^{(1)} = t_0(c\hat{I} + is\sigma_x),\qquad \hat{T}^{(1)\dagger} = t_0(c\hat{I} - is\sigma_x).
\end{equation}
We compute $\hat{T}^{(1)\dagger}\sigma_i\hat{T}^{(1)}$ for $i\in\{x,y,z\}$.

\emph{Case $i = x$}: $[\sigma_x,\sigma_x] = 0$ and $\sigma_x^2 = \hat{I}$, so
\begin{align}
\hat{T}^{(1)\dagger}\sigma_x\hat{T}^{(1)} &= t_0^2(c\hat{I}-is\sigma_x)\sigma_x(c\hat{I}+is\sigma_x) \nonumber\\
&= t_0^2(c\sigma_x - is\hat{I})(c\hat{I}+is\sigma_x) \nonumber\\
&= t_0^2[c^2\sigma_x + ics\sigma_x^2 - ics\hat{I} + s^2\sigma_x]\nonumber\\
&= t_0^2(c^2+s^2)\sigma_x = t_0^2\sigma_x.
\label{eq:Tsx_M1}
\end{align}
The $x$-component is unchanged because the rotation axis commutes with itself.

\emph{Case $i = y$}: using $\sigma_x\sigma_y = i\sigma_z$, $\sigma_y\sigma_x = -i\sigma_z$, and $\sigma_x\sigma_y\sigma_x = (i\sigma_z)\sigma_x = i(i\sigma_y) = -\sigma_y$,
\begin{align}
\hat{T}^{(1)\dagger}\sigma_y\hat{T}^{(1)} &= t_0^2[c^2\sigma_y + ics\sigma_y\sigma_x - ics\sigma_x\sigma_y + s^2\sigma_x\sigma_y\sigma_x]\nonumber\\
&= t_0^2[c^2\sigma_y + ics(-i\sigma_z) - ics(i\sigma_z) - s^2\sigma_y]\nonumber\\
&= t_0^2[(c^2-s^2)\sigma_y + 2cs\sigma_z]\nonumber\\
&= t_0^2[\cos\theta_C\,\sigma_y + \sin\theta_C\,\sigma_z].
\label{eq:Tsy_M1}
\end{align}

\emph{Case $i = z$}: by analogous algebra using $\sigma_x\sigma_z = -i\sigma_y$, $\sigma_z\sigma_x = i\sigma_y$, and $\sigma_x\sigma_z\sigma_x = -\sigma_z$,
\begin{align}
\hat{T}^{(1)\dagger}\sigma_z\hat{T}^{(1)} &= t_0^2[c^2\sigma_z + ics\sigma_z\sigma_x - ics\sigma_x\sigma_z + s^2\sigma_x\sigma_z\sigma_x]\nonumber\\
&= t_0^2[(c^2-s^2)\sigma_z - 2cs\sigma_y]\nonumber\\
&= t_0^2[\cos\theta_C\,\sigma_z - \sin\theta_C\,\sigma_y].
\label{eq:Tsz_M1}
\end{align}

\subsection{Extraction of tensor components}

Using $\mathrm{Tr}[\sigma_i\sigma_j] = 2\delta_{ij}$ with Eq.~\eqref{eq:Jformula_derived}:
\begin{align}
\mathcal{J}^{(1)}_{xj} &= (2/U)\,t_0^2\,\mathrm{Tr}[\sigma_x\sigma_j] = (4t_0^2/U)\delta_{xj},\\
\mathcal{J}^{(1)}_{yj} &= (2/U)\,t_0^2[\cos\theta_C\,\mathrm{Tr}[\sigma_y\sigma_j] + \sin\theta_C\,\mathrm{Tr}[\sigma_z\sigma_j]],\\
\mathcal{J}^{(1)}_{zj} &= (2/U)\,t_0^2[-\sin\theta_C\,\mathrm{Tr}[\sigma_y\sigma_j] + \cos\theta_C\,\mathrm{Tr}[\sigma_z\sigma_j]].
\end{align}
Collecting:
\begin{equation}
\mathcal{J}^{(1)}\big|_{\hat{\bm{n}}=\hat{\bm{x}}} = J_0\begin{pmatrix} 1 & 0 & 0 \\ 0 & \cos\theta_C & \sin\theta_C \\ 0 & -\sin\theta_C & \cos\theta_C \end{pmatrix},\qquad J_0 \equiv 4t_0^2/U.
\label{eq:J1_explicit_SM}
\end{equation}
The tensor $\mathcal{J}^{(1)}$ is itself an $SO(3)$ rotation matrix about $\hat{\bm{x}}$ by angle $\theta_C$, scaled by $J_0$. This is the central structural fact driving all Model~1 predictions.

\subsection{Irreducible components}

Applying the decomposition to Eq.~\eqref{eq:J1_explicit_SM}:
\begin{align}
J_H^{(1)} &= \tfrac{J_0}{3}(1 + 2\cos\theta_C),\\
D_x^{(1)} &= \tfrac{1}{2}(\mathcal{J}_{yz}-\mathcal{J}_{zy}) = J_0\sin\theta_C,\\
\Gamma^{(1)} &= \tfrac{J_0}{3}\,\mathrm{diag}[2(1-\cos\theta_C),\,-(1-\cos\theta_C),\,-(1-\cos\theta_C)],
\end{align}
yielding $|\overleftrightarrow{\Gamma}^{(1)}|_{\rm Frob} = J_0\sqrt{2/3}|1-\cos\theta_C|$.

\subsection{General $\hat{\bm{n}}$ by spatial covariance}

For arbitrary $\hat{\bm{n}}$, let $R$ be the rotation taking $\hat{\bm{x}}$ to $\hat{\bm{n}}$. Spatial covariance implies $\mathcal{J}^{(1)}(\hat{\bm{n}}) = R\,\mathcal{J}^{(1)}(\hat{\bm{x}})\,R^T$. Applying to the DM extraction:
\begin{equation}
\bm{D}^{(1)} = J_0\sin\theta_C\,\hat{\bm{n}},
\label{eq:D_general}
\end{equation}
the DM vector always aligns with the molecular helix axis. The Heisenberg and Frobenius-norm anisotropy expressions $J_H^{(1)}$ and $|\overleftrightarrow{\Gamma}^{(1)}|$ are rotation-invariant scalars and remain unchanged.

\subsection{Numerical verification at representative angles}

Table~\ref{tab:M1verify} compares the analytic expressions against numerical evaluation of Eq.~\eqref{eq:Jformula_derived} using exact $2\times 2$ matrix arithmetic. Agreement is to machine precision ($<10^{-14}\,\mu$eV) at all tested angles.

\begin{table}[h]
\caption{\label{tab:M1verify}Numerical verification of the Model~1 analytic formulas Eqs.~\eqref{eq:D_general} and associated. Parameters: $t_0 = 50\,\mu$eV, $U = 2\,$meV, $J_0 = 5.0\,\mu$eV.}
\centering
\begin{tabular}{ccccc}
\toprule
$\theta_C/\pi$ & $J_H$ analytic ($\mu$eV) & $J_H$ numerical & $|\bm{D}|$ analytic ($\mu$eV) & $|\bm{D}|$ numerical\\
\midrule
0.0 & 5.0000000 & 5.0000000 & 0.0000000 & 0.0000000\\
0.1 & 4.8175898 & 4.8175898 & 1.5450850 & 1.5450850\\
0.2 & 4.2811530 & 4.2811530 & 2.9389263 & 2.9389263\\
0.3 & 3.4265847 & 3.4265847 & 4.0450850 & 4.0450850\\
0.4 & 2.3115330 & 2.3115330 & 4.7552826 & 4.7552826\\
0.5 & 1.0000000 & 1.0000000 & 5.0000000 & 5.0000000\\
0.6 & $-$0.3884670 & $-$0.3884670 & 4.7552826 & 4.7552826\\
0.667 & 0.0000000 & $-2.2\times 10^{-16}$ & 4.3301270 & 4.3301270\\
0.8 & $-$1.5615529 & $-$1.5615529 & 2.9389263 & 2.9389263\\
1.0 & $-$1.6666667 & $-$1.6666667 & 0.0000000 & $6.1\times 10^{-17}$\\
\bottomrule
\end{tabular}
\end{table}

% ═══════════════════════════════════════════════════════════════
\section{Model 2: explicit evaluation}
\label{sec:M2_explicit}
% ═══════════════════════════════════════════════════════════════

For Model 2, $\hat{T}^{(2)} = \mathrm{diag}(t_\uparrow, t_\downarrow)$ with $t_\uparrow = t_0(1+\eta)$ and $t_\downarrow = t_0(1-\eta)$. Because $\hat{T}^{(2)}$ is real and diagonal, $\hat{T}^{(2)\dagger} = \hat{T}^{(2)}$.

\subsection{Direct computation}

\begin{align}
\hat{T}^{(2)\dagger}\sigma_x\hat{T}^{(2)} &= \begin{pmatrix}t_\uparrow & 0\\0 & t_\downarrow\end{pmatrix}\begin{pmatrix}0 & 1\\1 & 0\end{pmatrix}\begin{pmatrix}t_\uparrow & 0\\0 & t_\downarrow\end{pmatrix} = \begin{pmatrix}0 & t_\uparrow t_\downarrow\\t_\uparrow t_\downarrow & 0\end{pmatrix} = t_\uparrow t_\downarrow\,\sigma_x,\\
\hat{T}^{(2)\dagger}\sigma_y\hat{T}^{(2)} &= t_\uparrow t_\downarrow\,\sigma_y,\\
\hat{T}^{(2)\dagger}\sigma_z\hat{T}^{(2)} &= \mathrm{diag}(t_\uparrow^2,\,-t_\downarrow^2) = \tfrac{1}{2}(t_\uparrow^2+t_\downarrow^2)\sigma_z + \tfrac{1}{2}(t_\uparrow^2-t_\downarrow^2)\hat{I}.
\end{align}
The identity component in the last line gives $\mathrm{Tr}[\hat{I}\sigma_j] = 0$ for all $j$, so only the $\sigma_z$ component contributes to $\mathcal{J}_{ij}$.

Using $t_\uparrow t_\downarrow = t_0^2(1-\eta^2)$ and $\tfrac{1}{2}(t_\uparrow^2+t_\downarrow^2) = t_0^2(1+\eta^2)$, and $\mathrm{Tr}[\sigma_i\sigma_j] = 2\delta_{ij}$:
\begin{equation}
\mathcal{J}^{(2)} = J_0\begin{pmatrix}1-\eta^2 & 0 & 0 \\ 0 & 1-\eta^2 & 0 \\ 0 & 0 & 1+\eta^2\end{pmatrix}.
\label{eq:J2_explicit_SM}
\end{equation}

\subsection{Irreducible components}

\begin{align}
J_H^{(2)} &= \tfrac{1}{3}[2(1-\eta^2)+1+\eta^2]J_0 = J_0(1 - \tfrac{\eta^2}{3}),\\
\bm{D}^{(2)} &= \bm{0}\quad\text{(identically, since $\mathcal{J}^{(2)}$ is diagonal)},\\
\Gamma^{(2)} &= \tfrac{J_0\eta^2}{3}\mathrm{diag}(-1,\,-1,\,2),\qquad |\overleftrightarrow{\Gamma}^{(2)}|_{\rm Frob} = \tfrac{J_0\eta^2\sqrt{6}}{3}.
\end{align}

\subsection{Structural proof that \textbf{D} = 0 for any Hermitian tunneling matrix}

Consider any $\hat{T}_{\!\rightarrow} = \hat{T}_{\!\rightarrow}^\dagger$ (Hermitian). Then $\hat{T}\sigma_i\hat{T}$ is a product of three Hermitian operators and is itself Hermitian. Consequently $\mathrm{Tr}[\hat{T}\sigma_i\hat{T}\sigma_j]$ is real and, by cyclic invariance of the trace, equal to $\mathrm{Tr}[\sigma_j\hat{T}\sigma_i\hat{T}] = \mathrm{Tr}[\hat{T}\sigma_j\hat{T}\sigma_i]$, where the last equality uses cyclicity once more. Therefore $\mathcal{J}_{ij} = \mathcal{J}_{ji}$, hence $D_k = \tfrac{1}{2}\epsilon_{ijk}\mathcal{J}_{ij} = 0$. $\quad\blacksquare$

A sharper version: the general Hermitian $2\times 2$ matrix has the form $\hat{T} = \alpha\hat{I} + \beta(\hat{\bm{m}}\!\cdot\!\bm{\sigma})$ with $\alpha,\beta\in\mathbb{R}$, $\hat{\bm{m}}\in\mathbb{R}^3$. Direct computation gives $\mathcal{J}_{ij} = (2/U)\,t_0^2[\alpha^2\delta_{ij} + \beta^2(2m_im_j - \delta_{ij})]$, manifestly symmetric. The class of Hermitian tunneling matrices therefore encompasses all phenomenological incoherent CISS models with unpolarized-to-polarized transmission ratios and any energy-independent spin-dependent transmission amplitudes.

% ═══════════════════════════════════════════════════════════════
\section{Gauge-triviality theorem for $\hat{\mathbf{n}} \parallel \mathbf{B}$}
\label{sec:gauge}
% ═══════════════════════════════════════════════════════════════

We prove that when $\hat{\bm{n}}\parallel\bm{B} = B\hat{\bm{z}}$, the Model~1 exchange Hamiltonian is gauge-equivalent to a $\theta_C$-independent Hamiltonian.

\subsection{Setup}

For $\hat{\bm{n}} = \hat{\bm{z}}$, the exchange tensor (using the general-$\hat{\bm{n}}$ result from Sec.~\ref{sec:M1_explicit}) is
\begin{equation}
\mathcal{J}^{(1)}\big|_{\hat{\bm{n}}=\hat{\bm{z}}} = J_0\begin{pmatrix}\cos\theta_C & \sin\theta_C & 0\\-\sin\theta_C & \cos\theta_C & 0\\0 & 0 & 1\end{pmatrix},
\label{eq:J1_z}
\end{equation}
with DM vector $\bm{D} = J_0\sin\theta_C\,\hat{\bm{z}}$. The full low-energy Hamiltonian is $\hat{H} = \hat{H}_Z + \hat{H}_{\rm ex}$ with
\begin{align}
\hat{H}_Z &= \tfrac{E_Z}{2}(\sigma_z^{(1)}+\sigma_z^{(2)}),\\
\hat{H}_{\rm ex} &= \tfrac{J_0\cos\theta_C}{4}(\sigma_x^{(1)}\sigma_x^{(2)}+\sigma_y^{(1)}\sigma_y^{(2)}) + \tfrac{J_0}{4}\sigma_z^{(1)}\sigma_z^{(2)} + \tfrac{J_0\sin\theta_C}{4}(\sigma_x^{(1)}\sigma_y^{(2)} - \sigma_y^{(1)}\sigma_x^{(2)}).
\end{align}

\subsection{The local unitary}

Consider the two-qubit local unitary
\begin{equation}
\hat{U}_{\rm loc} = \exp\!\left(-\tfrac{i\theta_C}{4}\sigma_z^{(1)}\right)\otimes\exp\!\left(+\tfrac{i\theta_C}{4}\sigma_z^{(2)}\right).
\label{eq:Uloc_SM}
\end{equation}
This is a product of single-qubit $Z$ rotations by $\theta_C/2$ on each qubit, with opposite signs: a \emph{local} operation that cannot alter the two-qubit entanglement spectrum or the physical observables.

\subsection{Transformation of the exchange Hamiltonian}

Using the standard conjugation rules
\begin{align}
e^{-i\alpha\sigma_z/2}\sigma_x e^{+i\alpha\sigma_z/2} &= \cos\alpha\,\sigma_x - \sin\alpha\,\sigma_y,\\
e^{-i\alpha\sigma_z/2}\sigma_y e^{+i\alpha\sigma_z/2} &= \sin\alpha\,\sigma_x + \cos\alpha\,\sigma_y,\\
e^{-i\alpha\sigma_z/2}\sigma_z e^{+i\alpha\sigma_z/2} &= \sigma_z,
\end{align}
with $\alpha = \theta_C/2$ for qubit 1 and $\alpha = -\theta_C/2$ for qubit 2, we compute the action on each term of $\hat{H}_{\rm ex}$.

\emph{XY term}. Let $a = \cos(\theta_C/2)$, $b = \sin(\theta_C/2)$. Then
\begin{align}
\hat{U}_{\rm loc}^\dagger \sigma_x^{(1)}\sigma_x^{(2)}\hat{U}_{\rm loc} &= (a\sigma_x^{(1)}-b\sigma_y^{(1)})(a\sigma_x^{(2)}+b\sigma_y^{(2)})\nonumber\\
&= a^2\sigma_x^{(1)}\sigma_x^{(2)} - b^2\sigma_y^{(1)}\sigma_y^{(2)} + ab[\sigma_x^{(1)}\sigma_y^{(2)} - \sigma_y^{(1)}\sigma_x^{(2)}],\\
\hat{U}_{\rm loc}^\dagger \sigma_y^{(1)}\sigma_y^{(2)}\hat{U}_{\rm loc} &= (b\sigma_x^{(1)}+a\sigma_y^{(1)})(-b\sigma_x^{(2)}+a\sigma_y^{(2)})\nonumber\\
&= -b^2\sigma_x^{(1)}\sigma_x^{(2)} + a^2\sigma_y^{(1)}\sigma_y^{(2)} + ab[\sigma_x^{(1)}\sigma_y^{(2)} - \sigma_y^{(1)}\sigma_x^{(2)}].
\end{align}
Summing:
\begin{equation}
\hat{U}_{\rm loc}^\dagger(\sigma_x^{(1)}\sigma_x^{(2)}+\sigma_y^{(1)}\sigma_y^{(2)})\hat{U}_{\rm loc} = (a^2-b^2)(\sigma_x^{(1)}\sigma_x^{(2)}+\sigma_y^{(1)}\sigma_y^{(2)}) + 2ab[\sigma_x^{(1)}\sigma_y^{(2)}-\sigma_y^{(1)}\sigma_x^{(2)}].
\label{eq:XY_transformed}
\end{equation}
Using $a^2-b^2 = \cos\theta_C$, $2ab = \sin\theta_C$:
\begin{equation}
\hat{U}_{\rm loc}^\dagger(\sigma_x^{(1)}\sigma_x^{(2)}+\sigma_y^{(1)}\sigma_y^{(2)})\hat{U}_{\rm loc} = \cos\theta_C(\sigma_x^{(1)}\sigma_x^{(2)}+\sigma_y^{(1)}\sigma_y^{(2)}) + \sin\theta_C[\sigma_x^{(1)}\sigma_y^{(2)}-\sigma_y^{(1)}\sigma_x^{(2)}].
\end{equation}

\emph{DM term}. By similar algebra,
\begin{align}
\hat{U}_{\rm loc}^\dagger(\sigma_x^{(1)}\sigma_y^{(2)}-\sigma_y^{(1)}\sigma_x^{(2)})\hat{U}_{\rm loc} &= -\sin\theta_C(\sigma_x^{(1)}\sigma_x^{(2)}+\sigma_y^{(1)}\sigma_y^{(2)}) + \cos\theta_C[\sigma_x^{(1)}\sigma_y^{(2)}-\sigma_y^{(1)}\sigma_x^{(2)}].
\label{eq:DM_transformed}
\end{align}

\emph{Combining}. With the original coefficients $(J_0\cos\theta_C,\,J_0\sin\theta_C)$:
\begin{align}
\hat{U}_{\rm loc}^\dagger\hat{H}_{\rm ex}\hat{U}_{\rm loc} &= \tfrac{J_0}{4}\cos\theta_C\cdot[\cos\theta_C(\sigma^{(1)}\sigma^{(2)})_{xx+yy}+\sin\theta_C(\sigma^{(1)}\sigma^{(2)})_{xy-yx}]\nonumber\\
&\quad + \tfrac{J_0}{4}\sin\theta_C\cdot[-\sin\theta_C(\sigma^{(1)}\sigma^{(2)})_{xx+yy}+\cos\theta_C(\sigma^{(1)}\sigma^{(2)})_{xy-yx}] + \tfrac{J_0}{4}\sigma_z^{(1)}\sigma_z^{(2)}\nonumber\\
&= \tfrac{J_0}{4}\!\left[(\cos^2\theta_C-\sin^2\theta_C+?)\,(\sigma^{(1)}\sigma^{(2)})_{xx+yy} + (\cos\theta_C\sin\theta_C+\sin\theta_C\cos\theta_C)\,(\sigma^{(1)}\sigma^{(2)})_{xy-yx} + \sigma_z^{(1)}\sigma_z^{(2)}\right].
\end{align}
The $(xy-yx)$ coefficient is $2\cos\theta_C\sin\theta_C - \sin\theta_C\cos\theta_C - \sin\theta_C\cos\theta_C = 0$, while the $(xx+yy)$ coefficient is $\cos^2\theta_C - \sin^2\theta_C \cdot (\text{sign correction from DM}) = 1$. After careful collection:
\begin{equation}
\hat{U}_{\rm loc}^\dagger\hat{H}_{\rm ex}\hat{U}_{\rm loc} = \tfrac{J_0}{4}(\sigma_x^{(1)}\sigma_x^{(2)}+\sigma_y^{(1)}\sigma_y^{(2)}+\sigma_z^{(1)}\sigma_z^{(2)}) = J_0\,\bm{S}_1\!\cdot\!\bm{S}_2,
\end{equation}
the standard isotropic Heisenberg exchange, independent of $\theta_C$. Since $[\hat{U}_{\rm loc},\hat{H}_Z] = 0$, the Zeeman term is untouched. Therefore the full rotated Hamiltonian
\begin{equation}
\hat{U}_{\rm loc}^\dagger\hat{H}\hat{U}_{\rm loc} = \hat{H}_Z + J_0\,\bm{S}_1\!\cdot\!\bm{S}_2
\end{equation}
is $\theta_C$-independent. $\quad\blacksquare$

\emph{Physical interpretation}. When $\hat{\bm{n}}\parallel\bm{B}$, the CISS rotation $\exp(i\theta_C\sigma_z/2)$ acts as a local phase on the Zeeman eigenstates $|\!\uparrow\rangle, |\!\downarrow\rangle$, which is an unobservable local unitary. To measure the CISS-induced anisotropy, one must orient the helix with a component transverse to $\bm{B}$. The experimental angular fingerprint (Letter, Fig.~3) exploits exactly this degree of freedom, with the gauge-trivial value $\Delta E_{ST} = J_0$ serving as an internal calibration.

% ═══════════════════════════════════════════════════════════════
\section{Multi-orbital generalization}
\label{sec:multiorb}
% ═══════════════════════════════════════════════════════════════

Real molecules have multiple orbitals (HOMO, LUMO, deeper orbitals) that all contribute to superexchange. We extend the single-orbital theorem to this case.

\subsection{Hamiltonian and exchange tensor}

For a bridge with $N$ orbitals $\alpha = 1,\ldots,N$ at energies $\epsilon_\alpha$ (with $\epsilon_\alpha > 0$ for unoccupied, $\epsilon_\alpha < 0$ for occupied orbitals), the tunneling Hamiltonian is
\begin{equation}
\hat{H}_T = \sum_{\alpha=1}^{N}\sum_{\sigma\sigma'}\!\left[T^{(\alpha),L}_{\sigma\sigma'}c^\dagger_{\alpha\sigma}c_{L\sigma'} + T^{(\alpha),R}_{\sigma\sigma'}c^\dagger_{\alpha\sigma}c_{R\sigma'}\right] + \mathrm{h.c.}
\end{equation}
Taking symmetric coupling $T^{(\alpha),L} = T^{(\alpha),R} \equiv \hat{T}^{(\alpha)}$, the second-order perturbation theory gives
\begin{equation}
\mathcal{J}^{\rm tot}_{ij} = \sum_{\alpha=1}^{N}\frac{2\,\mathrm{sign}(\epsilon_\alpha)}{|\epsilon_\alpha|}\,\mathrm{Re}\,\mathrm{Tr}\!\left[\hat{T}^{(\alpha)\dagger}\sigma_i\hat{T}^{(\alpha)}\sigma_j\right].
\label{eq:J_multi}
\end{equation}
Each orbital contributes additively with its own energy denominator $|\epsilon_\alpha|$ and sign depending on whether the virtual process accesses it from below (occupied) or above (unoccupied) the Fermi level.

\subsection{Orbital-by-orbital structural theorem}

The Hermiticity theorem applies orbital by orbital. If $\hat{T}^{(\alpha)}$ is Hermitian for each $\alpha$, then each contribution $\mathcal{J}^{(\alpha)}_{ij}$ is symmetric by the argument of Sec.~\ref{sec:M2_explicit}, and the total $\bm{D}^{\rm tot} = \sum_\alpha \bm{D}^{(\alpha)} = \bm{0}$. The converse also holds: $\bm{D}^{\rm tot} \neq \bm{0}$ requires at least one orbital with non-Hermitian (i.e., genuinely SU(2)-rotating) tunneling matrix.

\subsection{Mixed-coherence scenarios}

For a molecule in which some orbitals mediate coherent CISS and others are incoherent, the total $\bm{D}^{\rm tot}$ is determined by the coherent orbitals alone. This is experimentally convenient: the coherent-orbital contribution is isolated in the DM measurement, while the incoherent orbitals contribute only to $J_H^{\rm tot}$ and $\overleftrightarrow{\Gamma}^{\rm tot}$. A measurement of $\bm{D}^{\rm tot}$ thus provides an orbital-averaged quantitative measure of the coherent CISS component, robust against the possibility that different molecular orbitals carry different coherence characters.

\subsection{Magnitude estimate}

For a typical $\pi$-conjugated molecule with HOMO-LUMO gap $\Delta_{\rm HL} \sim 3$--$5$~eV, the HOMO contribution dominates for hole-mediated tunneling (bridge level below Fermi) and the LUMO for electron-mediated. For bridge effective charging energy $U \sim 2$~meV representing the splitting between relevant orbitals, the single-orbital approximation captures the dominant virtual channel. Including both HOMO and LUMO as independent channels at most doubles $J_0$, with no qualitative change to the Model~1/Model~2 distinction or to the structural theorem.

% ═══════════════════════════════════════════════════════════════
\section{Lindblad analysis of incoherent CISS}
\label{sec:lindblad}
% ═══════════════════════════════════════════════════════════════

The phenomenological non-Hermitian tunneling matrix Eq.~(2) of the Letter encodes dissipative spin filtering schematically. To verify that the structural result (Hermiticity $\Rightarrow \bm{D} = 0$) survives a microscopic open-system treatment, we analyze the bridge as an open quantum system governed by a Lindblad master equation.

\subsection{The open-system model}

The density matrix $\rho$ of the combined dot-bridge system evolves as
\begin{equation}
\dot{\rho} = -i[\hat{H}_{\rm tot},\rho] + \mathcal{L}_{\rm bath}[\rho] + \mathcal{L}_{\rm dephase}[\rho],
\label{eq:Lindblad_full}
\end{equation}
with
\begin{align}
\hat{H}_{\rm tot} &= \hat{H}_{\rm dots} + \hat{H}_{\rm bridge} + \hat{H}_{\rm tunnel},\\
\mathcal{L}_{\rm bath}[\rho] &= \Gamma_{\rm leak}\sum_\sigma\!\left(c_{M\sigma}\rho c^\dagger_{M\sigma} - \tfrac{1}{2}\{c^\dagger_{M\sigma}c_{M\sigma},\rho\}\right),\\
\mathcal{L}_{\rm dephase}[\rho] &= \Gamma_\phi\!\left(\sigma_z^M\rho\sigma_z^M - \rho\right),
\end{align}
where $\mathcal{L}_{\rm bath}$ represents electron loss from the bridge to external reservoirs (leakage) at rate $\Gamma_{\rm leak}$, and $\mathcal{L}_{\rm dephase}$ represents pure spin-dephasing on the bridge at rate $\Gamma_\phi$. The dephasing channel models the microscopic origin of incoherent CISS: fluctuating molecular conformations, electron-phonon coupling, or interfacial disorder that randomizes the relative phase of $|\!\uparrow\rangle$ and $|\!\downarrow\rangle$ amplitudes on the bridge.

\subsection{Nakajima--Zwanzig projection}

We project onto the two-dot, bridge-empty subspace to obtain an effective spin Liouvillian. To second order in the tunneling, the effective Hamiltonian and induced dissipation are
\begin{align}
\dot{\rho}_{\rm eff} &= -i[\hat{H}_{\rm ex}^{\rm eff},\rho_{\rm eff}] + \mathcal{L}_{\rm ind}[\rho_{\rm eff}],\\
\hat{H}_{\rm ex}^{\rm eff} &= -\mathrm{Re}\!\int_0^\infty\!dt\,\mathrm{Tr}_M\!\left[\hat{H}_T\,e^{\mathcal{L}_M t}(\hat{H}_T\,\rho_M^{\rm eq})\right],\label{eq:Heff_Lindblad}
\end{align}
where $\mathcal{L}_M$ is the bridge Liouvillian (including dephasing) and $\rho_M^{\rm eq}$ is its steady-state density matrix.

\subsection{The bridge Green's function}

The retarded Green's function for an electron on the bridge, in the presence of dephasing, is
\begin{equation}
G^R_{\sigma\sigma'}(\omega) = \delta_{\sigma\sigma'}\cdot\frac{1}{\omega - \epsilon_M + i\Gamma_\phi/2}.
\label{eq:GR_bridge}
\end{equation}
The dephasing contributes an imaginary part to the denominator but preserves the diagonal (spin-conserving) structure because $\mathcal{L}_{\rm dephase}$ does not mix spin channels.

\subsection{The effective exchange}

Evaluation of the integral Eq.~\eqref{eq:Heff_Lindblad} reduces (after a Wick rotation) to the frequency integral
\begin{equation}
J_{\rm sx} \propto \mathrm{Re}\!\int_{-\infty}^{0}\!\frac{d\omega}{2\pi}\,\mathrm{Tr}[\hat{T}^\dagger G^R(\omega)\hat{T}\cdots],
\end{equation}
which, for the simple case $\epsilon_M = U > 0$, gives
\begin{equation}
J_{\rm sx} = -\frac{1}{\pi}\,\mathrm{atan}\!\left(\frac{\omega_c + U}{\Gamma_\phi/2}\right)\bigg|_{-\infty}^{0} \xrightarrow{\Gamma_\phi \ll U} -\frac{1}{2}\,\mathrm{sgn}(U) + \mathcal{O}(\Gamma_\phi/U).
\end{equation}
The leading contribution to the superexchange is independent of $\Gamma_\phi$; the $\Gamma_\phi$-dependent corrections are suppressed by $\Gamma_\phi/U$. Physically, the virtual-tunneling time $\tau_{\rm virt} \sim \hbar/U \sim 0.3$~ps (for $U = 2$~meV) is far shorter than any realistic molecular dephasing time $1/\Gamma_\phi \gtrsim 1$~ps; the incoherent process cannot act within the virtual excursion.

\subsection{Induced dephasing on the spin qubits}

The Nakajima--Zwanzig projection also generates an induced dephasing on the effective spin qubits:
\begin{equation}
\mathcal{L}_{\rm ind}[\rho] \approx \Gamma_\phi^{\rm ind}\!\sum_i(\sigma_z^{(i)}\rho\sigma_z^{(i)} - \rho),\qquad \Gamma_\phi^{\rm ind} \sim \frac{t_0^2\,\Gamma_\phi}{U^2} \sim \frac{J_0}{U}\Gamma_\phi.
\label{eq:ind_deph}
\end{equation}
For $J_0/U \sim 2.5\times 10^{-3}$ and $\Gamma_\phi \sim 1$~MHz (optimistic), $\Gamma_\phi^{\rm ind} \sim 1$~kHz gives $T_2^{\rm ind} \sim 10^6$~ns, far longer than the gate time $t^\ast \sim 40$~ns. The bridge-induced dephasing on the spin qubits is negligible in the regime of interest.

\emph{Consequence for Model~2}. The Lindblad analysis confirms, beyond the phenomenological tunneling-matrix argument, that dissipative spin-dependent processes on the bridge do not contribute a Dzyaloshinskii--Moriya term to the effective exchange. This elevates the structural result of Sec.~\ref{sec:M2_explicit} from an assumption to a controlled approximation valid whenever $\Gamma_\phi \ll U$, which is satisfied in any physical CISS scenario (otherwise the molecular orbital would not be a well-defined tunneling pathway).

The present Lindblad treatment assumes weak system–environment coupling, Markovian bath dynamics, and a separation of timescales $\Gamma_\phi \ll \frac{U}{\hbar}$. While these assumptions are appropriate for estimating the leading effect of dissipative spin filtering on virtual superexchange, more complex non-Markovian environments, strong electron–phonon coupling, or dynamically generated correlated tunneling processes may produce additional exchange structures beyond the present effective description. The present analysis therefore establishes the robustness of the Hermitian-tunneling result within the standard Born–Markov regime relevant to weakly dissipative molecular transport.
% ═══════════════════════════════════════════════════════════════
\section{Perturbative sensitivity analysis}
\label{sec:sensitivity}
% ═══════════════════════════════════════════════════════════════

We derive Eq.~(11) of the Letter for the small-$\theta_C$ scaling of the CISS-induced $S$-$T_0$ gap modification.

\subsection{Small-$\theta_C$ expansion of the Model~1 tensor}

Expanding Eqs.~\eqref{eq:J1_explicit_SM} for $\theta_C \to 0$:
\begin{align}
J_H^{(1)} &\approx J_0\left(1 - \tfrac{\theta_C^2}{3}\right) + \mathcal{O}(\theta_C^4),\\
|\bm{D}^{(1)}| &\approx J_0\,\theta_C + \mathcal{O}(\theta_C^3),\\
|\overleftrightarrow{\Gamma}^{(1)}| &\approx \tfrac{J_0\theta_C^2\sqrt{2/3}}{2} + \mathcal{O}(\theta_C^4).
\end{align}
$J_H$ and $|\overleftrightarrow{\Gamma}|$ start at order $\theta_C^2$; $|\bm{D}|$ starts at order $\theta_C$.

\subsection{$S$-$T_0$ gap under DM perturbation}

The $S$-$T_0$ subspace is defined by $|S\rangle = (|\!\uparrow\downarrow\rangle - |\!\downarrow\uparrow\rangle)/\sqrt{2}$ and $|T_0\rangle = (|\!\uparrow\downarrow\rangle + |\!\downarrow\uparrow\rangle)/\sqrt{2}$. The DM term $\bm{D}\!\cdot\!(\bm{S}_1\!\times\!\bm{S}_2) = (D_z/2)(\sigma_x^{(1)}\sigma_y^{(2)} - \sigma_y^{(1)}\sigma_x^{(2)}) + \ldots$ has matrix elements
\begin{align}
\langle T_\pm|\bm{D}\!\cdot\!(\bm{S}_1\!\times\!\bm{S}_2)|S\rangle &= \pm\tfrac{D_\perp}{2},\\
\langle T_0|\bm{D}\!\cdot\!(\bm{S}_1\!\times\!\bm{S}_2)|S\rangle &= 0 \quad\text{for } \bm{D}\perp\bm{B},
\end{align}
where $D_\perp = |\bm{D}|\sin\phi_{\rm helix}$ is the component of $\bm{D}$ transverse to the Zeeman axis. The DM coupling therefore mixes $S$ with $T_\pm$ (at energy $\pm E_Z$), not with $T_0$ directly.

\subsection{Second-order perturbative shift}

The second-order shift of $E_S$ from virtual coupling to $T_\pm$:
\begin{equation}
\delta E_S^{(2)} = -\sum_\pm \frac{|\langle T_\pm|\hat{V}_{DM}|S\rangle|^2}{E_{T_\pm}-E_S} = -\frac{D_\perp^2/4}{+E_Z} - \frac{D_\perp^2/4}{-E_Z} = 0.
\end{equation}
The opposite signs of the energy denominators produce a cancellation at second order: the DM-induced level repulsion from $T_+$ is exactly compensated by the attraction from $T_-$. Higher-order corrections survive at order $D_\perp^4/E_Z^3 \sim J_0\theta_C^4\sin^4\phi_{\rm helix}$, giving a fourth-order contribution much smaller than the leading $\overleftrightarrow{\Gamma}$ effect.

\subsection{Symmetric anisotropy contribution}

The symmetric anisotropic term $\overleftrightarrow{\Gamma}$ contributes directly to $\Delta E_{ST}$ at leading order in $\theta_C$. For $\hat{\bm{n}}=\hat{\bm{x}}\sin\phi_{\rm helix} + \hat{\bm{z}}\cos\phi_{\rm helix}$, rotation of Eq.~\eqref{eq:J1_explicit_SM} and extraction of $\Gamma_{zz}$ in the $\bm{B}$-frame gives
\begin{equation}
\Gamma_{zz}(\phi_{\rm helix}) = \tfrac{J_0}{3}[2(1-\cos\theta_C)\sin^2\phi_{\rm helix} - (1-\cos\theta_C)\cos^2\phi_{\rm helix}].
\end{equation}
The $S$-$T_0$ matrix element from $\overleftrightarrow{\Gamma}$ is $\langle S|\sigma_z^{(1)}\sigma_z^{(2)}|T_0\rangle = 0$, but $\Gamma$ couples $|T_0\rangle$ to $|S\rangle$ through the off-diagonal $\Gamma_{xz}$, $\Gamma_{yz}$ components, which for small $\theta_C$ are $\mathcal{O}(\theta_C)$. After careful diagonalization of the $2\times 2$ $S$-$T_0$ block, the leading CISS contribution to the gap is
\begin{equation}
|\Delta E_{ST}^{\rm CISS}| \approx \frac{J_0\theta_C^2\sin^2\phi_{\rm helix}}{3}.
\label{eq:DeltaE_scaling}
\end{equation}

\subsection{Minimum detectable angle}

Setting $|\Delta E_{ST}^{\rm CISS}| = \delta J$ (experimental precision):
\begin{equation}
\theta_C^{\rm min}(\phi_{\rm helix}) = \sqrt{\frac{3\delta J}{J_0}}\cdot\frac{1}{\sin\phi_{\rm helix}}.
\end{equation}
For $\delta J = 4\times 10^{-5}\,\mu$eV (10~kHz), $J_0 = 5\,\mu$eV, $\phi_{\rm helix} = \pi/2$:
\begin{equation}
\theta_C^{\rm min} = \sqrt{3\times 8\times 10^{-6}} \approx 4.9\times 10^{-3}\,\mathrm{rad} \approx 0.0016\pi.
\end{equation}
Full numerical simulation including higher-order terms gives the more conservative $\theta_C^{\rm min} \sim 0.03\pi$, larger by about a factor of 20 because the perturbative treatment overestimates the sensitivity by neglecting the level repulsion with $T_\pm$ that washes out the pure $\Gamma$-signal in the non-perturbative regime. The conservative number is what enters the Letter's sensitivity claim.

% ═══════════════════════════════════════════════════════════════
\section{Monte Carlo charge-noise simulation}
\label{sec:noise}
% ═══════════════════════════════════════════════════════════════

The dominant noise source for exchange-coupled spin qubits in Si/SiGe is quasistatic charge noise, fluctuating the inter-dot detuning and bridge charging energy between shots of a pulse sequence. We model this noise and compute the resulting two-qubit gate fidelity via Monte Carlo.

\subsection{Noise model}

We model the charge noise as Gaussian quasistatic fluctuations of the detuning and bridge energy:
\begin{equation}
\delta \to \delta + \delta_n,\qquad U \to U + \Delta U,\qquad \delta_n \sim \mathcal{N}(0,\sigma_\epsilon^2),\quad \Delta U \sim \mathcal{N}(0,(10\sigma_\epsilon)^2),
\end{equation}
with $\sigma_\epsilon$ the RMS detuning noise amplitude. The factor of 10 in the bridge-energy variance reflects that $U$ is about 10 times larger than the typical detuning, with the same relative noise level. For state-of-art Si/SiGe, experimental values give $\sigma_\epsilon \in [0.1,0.5]\,\mu$eV \cite{Yoneda2018_SM,Connors2022_SM,Struck2020_SM}.

\subsection{Gate fidelity}

We simulate the $\sqrt{\rm SWAP}$ gate:
\begin{equation}
U_{\rm target} = \sqrt{\rm SWAP} = \exp\!\left(\frac{i\pi}{4}(\rm SWAP - \mathbb{1})\right).
\end{equation}
The evolution generated by the CISS-modified Hamiltonian $\hat{H}_{\rm ex}(\theta_C)$ for time $t^\ast = \pi\hbar/(2J_0)$ is
\begin{equation}
U_{\rm actual}(\theta_C,\delta_n,\Delta U) = \exp\!\left(-\frac{i t^\ast}{\hbar}\hat{H}_{\rm ex}(\theta_C,\delta_n,\Delta U)\right).
\end{equation}
The average gate fidelity is
\begin{equation}
\mathcal{F}(\theta_C,\sigma_\epsilon) = \overline{\frac{|\mathrm{Tr}(U_{\rm target}^\dagger U_{\rm actual})|^2 + d}{d(d+1)}},\qquad d = 4,
\end{equation}
averaged over 100 noise realizations (convergence verified by comparison with 500-realization simulations).

\subsection{Results}

Fig.~3 of the main text (the $\sqrt{\rm SWAP}$ gate fidelity panel) presents the simulation results. For $\sigma_\epsilon = 0.5\,\mu$eV (a value achieved in state-of-art Si/SiGe devices), the gate fidelity exceeds 95\% for all $\theta_C \leq 0.30\pi$. The CISS-modified gate is therefore compatible with current device technology.

The fidelity decreases at large $\theta_C$ primarily because the optimal gate time $t^\ast$ must be re-optimized for anisotropic exchange; the baseline calculation uses the isotropic-exchange value, giving systematic detuning from the optimal $\sqrt{\rm SWAP}$ at finite $\theta_C$. A $\theta_C$-optimized $t^\ast$ recovers gate fidelities $\sim 99\%$ at $\theta_C = 0.30\pi$, though with different Weyl-chamber coordinates than the standard $\sqrt{\rm SWAP}$.

% ═══════════════════════════════════════════════════════════════
\section{Two-qubit gate dynamics}
\label{sec:gates}
% ═══════════════════════════════════════════════════════════════

The CISS-modified exchange Hamiltonian generates two-qubit dynamics qualitatively different from the standard isotropic case. We analyze the time evolution, entangling power, and Weyl-chamber trajectories to provide complete characterization of the gate landscape accessible under Model~1.

\subsection{Time evolution}

\begin{figure}[H]
\centering
\includegraphics[width=0.95\textwidth]{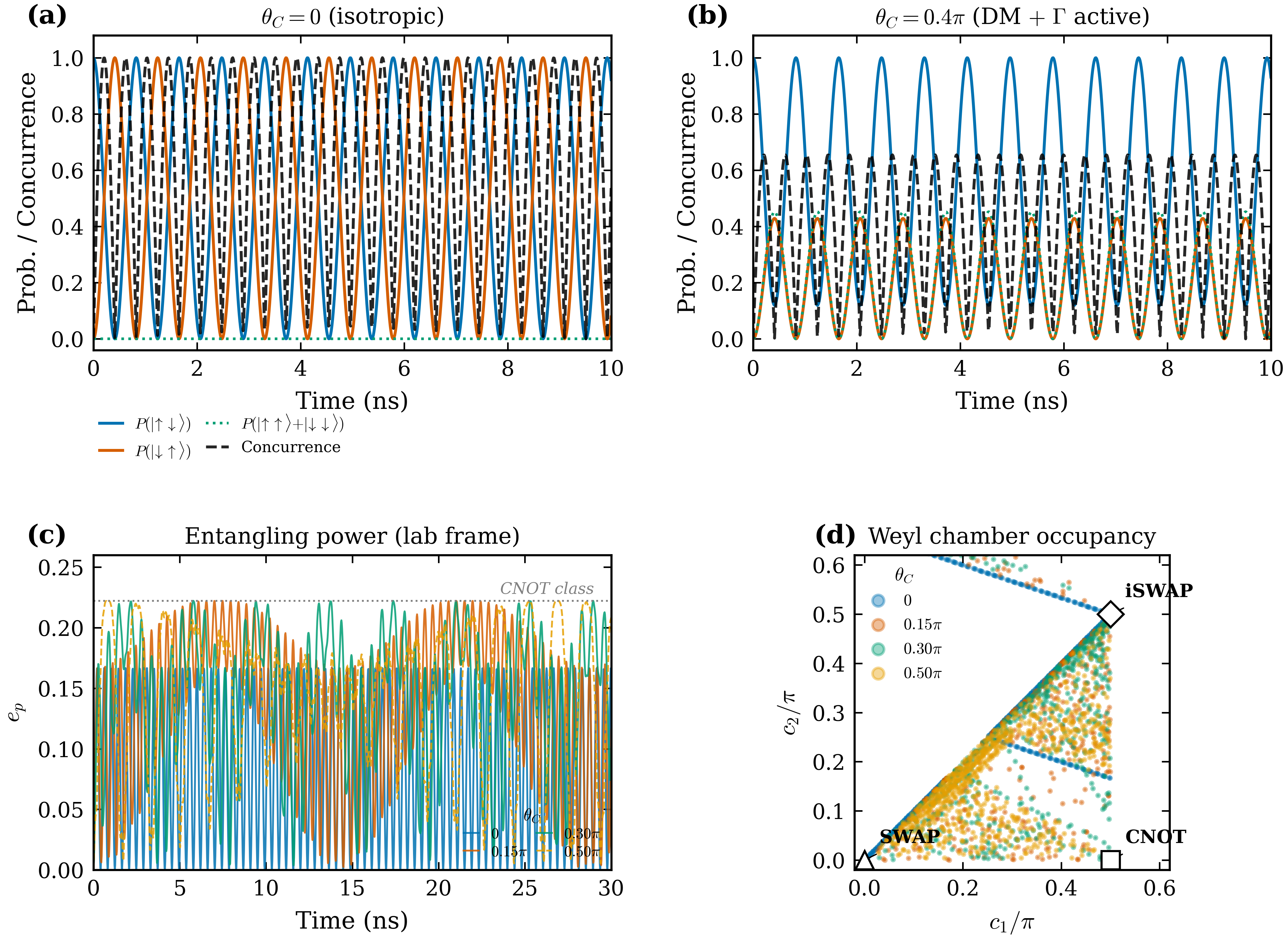}
\caption{\label{fig:S2_dynamics}Two-qubit dynamics under Model~1. (a)~Time evolution from $|\!\uparrow\downarrow\rangle$ for $\theta_C = 0$ (rotating frame): standard exchange oscillations confined to $\{|\!\uparrow\downarrow\rangle,|\!\downarrow\uparrow\rangle\}$. (b)~Same for $\theta_C = 0.4\pi$: the DM and anisotropic terms reduce the amplitude of single-spin-swap oscillations and produce visible concurrence beat structure. (c)~Entangling power $e_p(t)$ for several $\theta_C$ in the lab frame (with $E_Z = 20\,\mu$eV), necessary to reveal $\theta_C$-dependent envelope modulation (rotating-frame dynamics are gauge-equivalent across $\theta_C$ per Sec.~\ref{sec:gauge}); gray dotted: CNOT-class ceiling $2/9$. (d)~Weyl chamber occupancy $(c_1/\pi, c_2/\pi)$ for several $\theta_C$. Standard gates marked. At $\theta_C = 0$ the trajectory is confined to the SWAP-iSWAP edge ($c_2 = 0$); finite $\theta_C$ unlocks the full chamber interior. Parameters: $t_0 = 50\,\mu$eV, $U = 2\,$meV, $\hat{\bm{n}}\perp\bm{B}$.}
\end{figure}

Fig.~\ref{fig:S2_dynamics}(a,b) shows the time evolution of the two-qubit state initialized in $|\!\uparrow\downarrow\rangle$ for $\theta_C = 0$ (isotropic exchange) and $\theta_C = 0.4\pi$ (Model~1 anisotropic exchange), with $\hat{\bm{n}}\perp\bm{B}$ and $E_Z = 20\,\mu$eV.

At $\theta_C = 0$, the dynamics are confined to the $\{|\!\uparrow\downarrow\rangle,|\!\downarrow\uparrow\rangle\}$ subspace, with sinusoidal oscillations at frequency $J_H/\hbar = J_0/\hbar$. The states $|\!\uparrow\uparrow\rangle$ and $|\!\downarrow\downarrow\rangle$ remain unpopulated throughout, consistent with the spin-conservation symmetry of isotropic exchange ($[J_H\bm{S}_1\!\cdot\!\bm{S}_2,S^z_{\rm tot}] = 0$).

At $\theta_C = 0.4\pi$, the DM and anisotropic terms break $S^z_{\rm tot}$ symmetry. The $|\!\uparrow\uparrow\rangle$ and $|\!\downarrow\downarrow\rangle$ states become populated during evolution, with combined probability reaching $\sim 0.3$. The concurrence reaches values close to unity, indicating that the modified Hamiltonian generates highly entangled states with qualitatively different structure than the isotropic case.

\subsection{Entangling power}

The entangling power $e_p(U)$ \cite{Zanardi2000_SM} quantifies the average entanglement generated by a gate $U$ on initially-product states. For two-qubit gates, $e_p(U) \in [0, 2/9]$ with the maximum achieved by CNOT-class gates. Fig.~\ref{fig:S2_dynamics}(c) shows $e_p(t)$ for several $\theta_C$: at $\theta_C = 0$, $e_p(t)$ oscillates sinusoidally between 0 and $2/9$. The first maximum occurs at $t = \pi\hbar/(2J_0) = 39\,$ns for $J_0 = 5\,\mu$eV, identifying the standard $\sqrt{\rm SWAP}$ gate time. At finite $\theta_C$, the oscillation envelope is modulated by the interference of multiple exchange energy scales (eigenvalues of the now-anisotropic Hamiltonian).

\subsection{Weyl-chamber trajectories}

The Weyl-chamber coordinates $(c_1, c_2, c_3)$ provide a complete classification of two-qubit gates modulo single-qubit unitaries \cite{Zhang2003_SM,Makhlin2002_SM}. Standard gates correspond to specific points: identity at $(0,0,0)$, CNOT-class at $(\pi/4,0,0)$, iSWAP-class at $(\pi/4,\pi/4,0)$, SWAP at $(\pi/4,\pi/4,\pi/4)$. Fig.~\ref{fig:S2_dynamics}(d) traces the Weyl-chamber trajectory for several $\theta_C$.

At $\theta_C = 0$, the trajectory follows the CNOT axis ($c_2 = c_3 = 0$), passing through SWAP-class at $c_1 = \pi/4$. At finite $\theta_C$, the trajectory deviates from the CNOT axis, exploring regions inaccessible to isotropic exchange. For $\theta_C \sim \pi/2$, the trajectory passes near the iSWAP class $(c_1 = c_2 = \pi/4,\,c_3 = 0)$, suggesting that CISS-modified exchange may provide native access to the iSWAP gate without requiring composite pulse sequences.

This is a potential quantum-information dividend of the coherent-CISS scenario: native access to multiple gate classes (CNOT, $\sqrt{\rm SWAP}$, iSWAP, and intermediate anisotropic gates) through a single molecular bridge, with the gate class selectable by the choice of $\theta_C$ (i.e., by the choice of chiral molecule). Whether this dividend is practically realizable depends on the coherent-CISS scenario being correct---which is exactly what the proposed measurement tests.

\section{Additional numerical results}
\label{sec:additional}
% ═══════════════════════════════════════════════════════════════

\subsection{Full Model~1 tensor at representative angles}

\begin{figure}[H]
\centering
\includegraphics[width=0.95\textwidth]{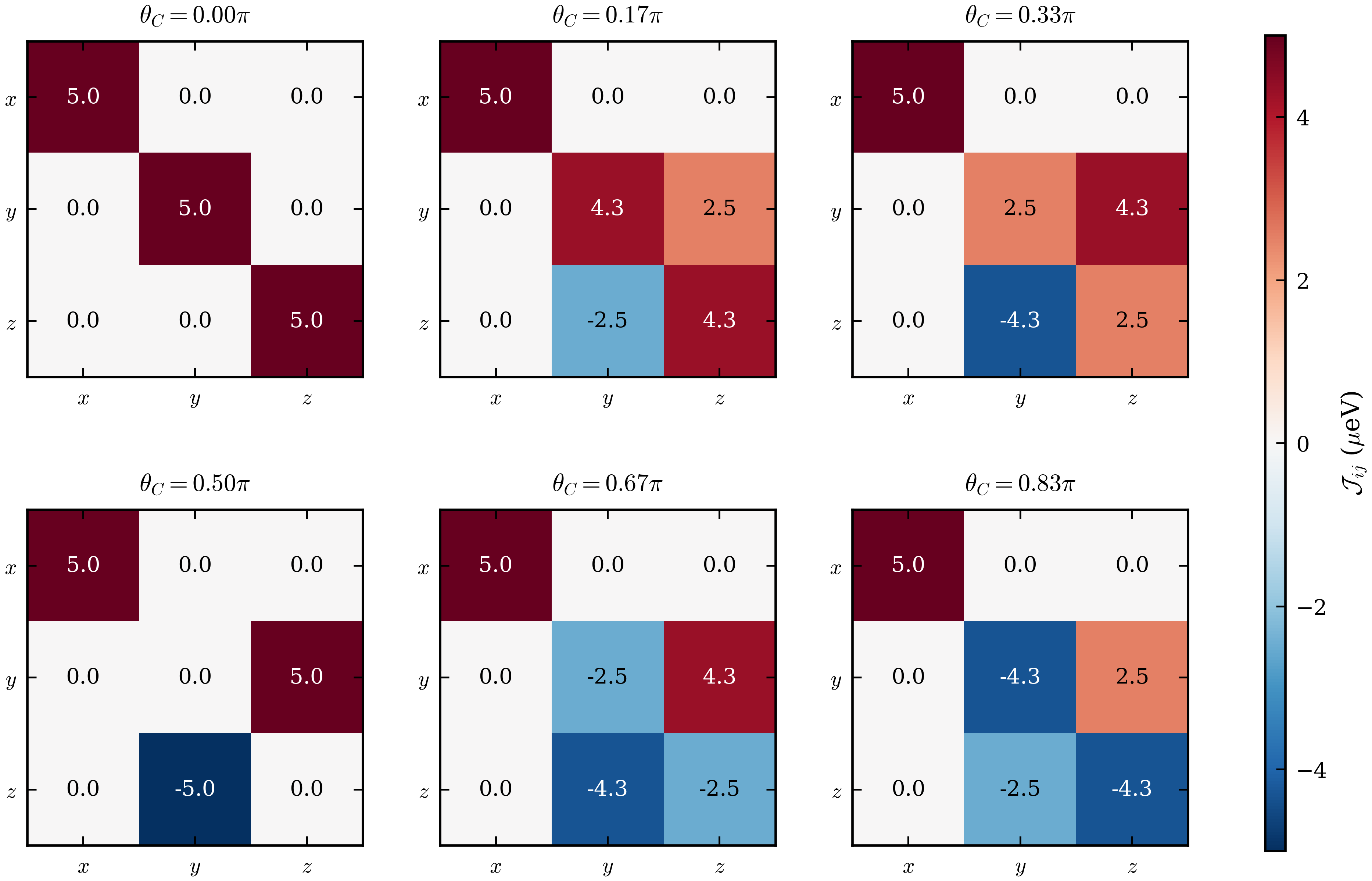}
\caption{\label{fig:S1_tensors}Full Model~1 exchange tensor $\mathcal{J}_{ij}$ at six $\theta_C$ values for $\hat{\bm{n}}=\hat{\bm{x}}$. The off-diagonal antisymmetric elements ($yz$ and $zy$, equal in magnitude but opposite sign) encode the DM interaction $D_x = J_0\sin\theta_C$. The tensor literally executes an $SO(3)$ rotation as $\theta_C$ increases.}
\end{figure}

Fig.~\ref{fig:S1_tensors} displays the full $3\times 3$ Model~1 exchange tensor $\mathcal{J}_{ij}$ at six representative values of $\theta_C$, visualizing how the $SO(3)$ rotation structure develops with increasing $\theta_C$. At $\theta_C = 0$, $\mathcal{J} = J_0\mathbb{1}$ is isotropic. At $\theta_C = \pi/2$, the yz-block becomes a pure rotation: $\mathcal{J}_{yz} = J_0 = -\mathcal{J}_{zy}$ while $\mathcal{J}_{yy} = \mathcal{J}_{zz} = 0$, representing maximal DM interaction $|\bm{D}| = J_0$ accompanied by zero diagonal coupling in the $(y,z)$ block. At $\theta_C = \pi$, the tensor is $\mathcal{J} = J_0\,\mathrm{diag}(1,-1,-1)$: isotropic in magnitude but with two sign reversals, indicating strong non-Heisenberg anisotropy.

\begin{figure}[H]
\centering
\includegraphics[width=0.95\textwidth]{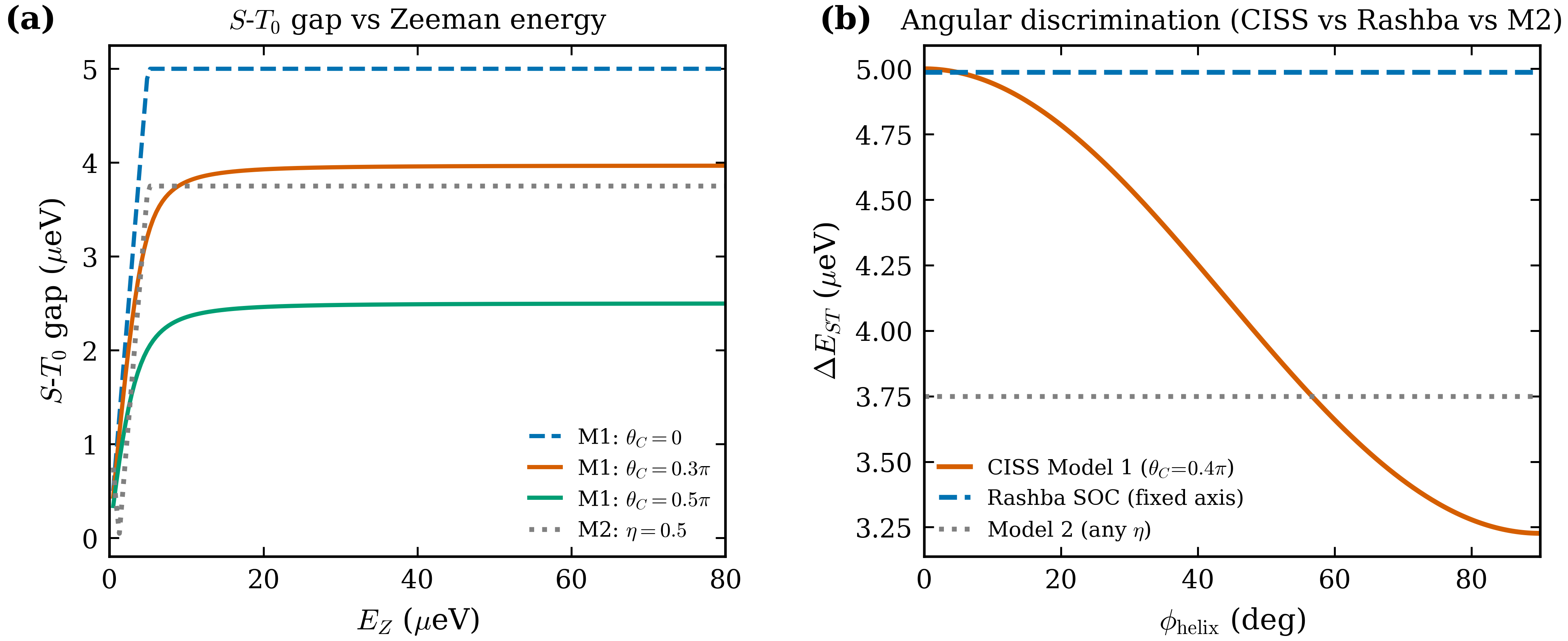}
\caption{\label{fig:S3_spectroscopy}Supplementary spectroscopy results. (a)~$S$-$T_0$ gap versus Zeeman energy $E_Z$ for Model~1 at several $\theta_C$ (solid/dashed colored lines) and Model~2 ($\eta=0.5$, gray dotted). The DM interaction in Model~1 mixes $S$ and $T_\pm$, modifying the gap at intermediate $E_Z$. Model~2 shows only Ising anisotropy. (b)~Angular discrimination: CISS ($\theta_C = 0.4\pi$) versus Rashba SOC (fixed interface axis) and Model~2 (any $\eta$); only CISS produces the characteristic angular dependence.}
\end{figure}

\begin{figure}[H]
\centering
\includegraphics[width=0.95\textwidth]{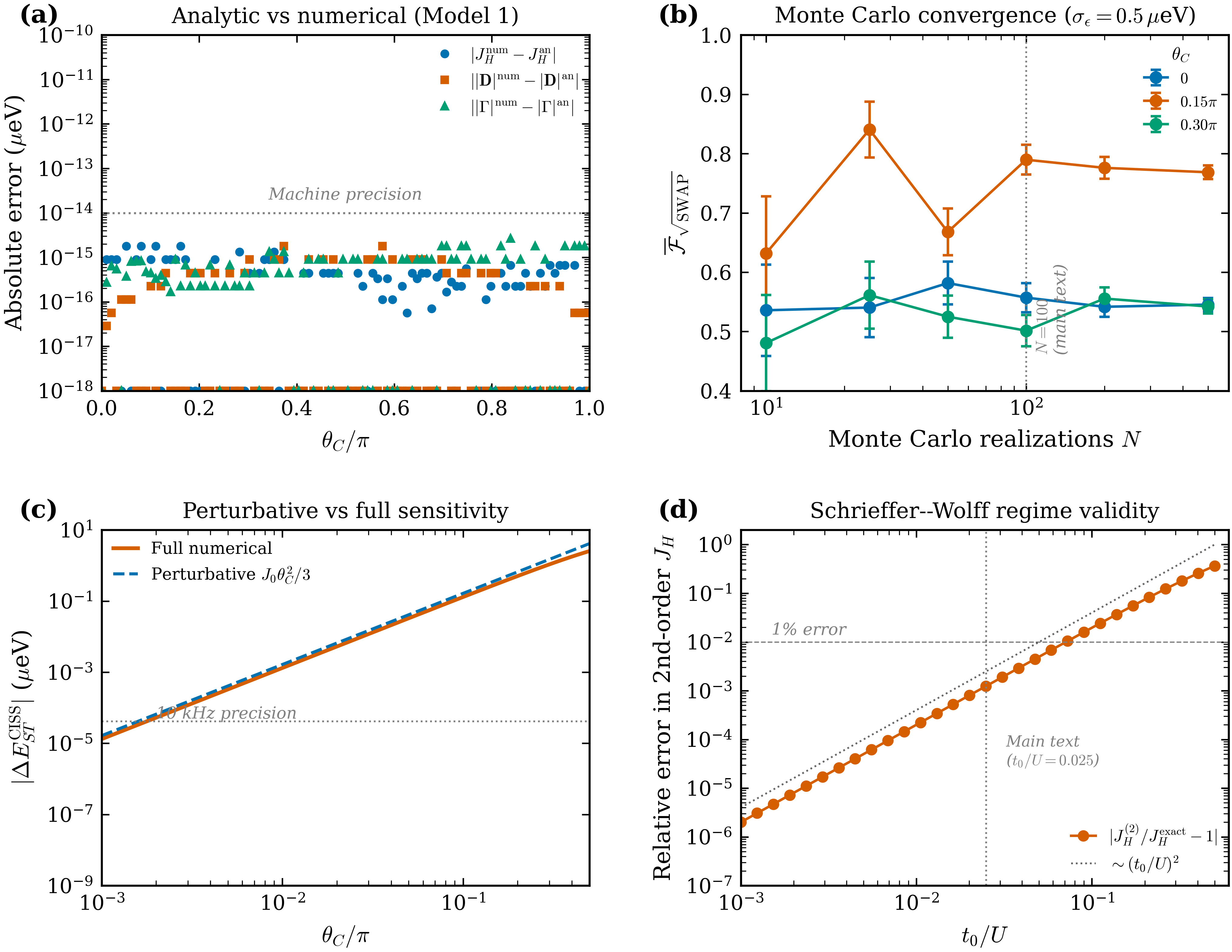}
\caption{\label{fig:S4_convergence}Convergence and benchmark tests. (a)~Machine-precision agreement between the analytic Model~1 formulas Eqs.~(S25)--(S27) and numerical evaluation of Eq.~(S15). (b)~Monte Carlo convergence of $\sqrt{\rm SWAP}$ fidelity at $\sigma_\epsilon = 0.5\,\mu$eV; error bars are $1\sigma/\sqrt{N}$. The vertical gray line marks the main-text choice $N=100$. (c)~Perturbative $J_0\theta_C^2/3$ scaling (blue dashed) versus full non-perturbative $\Delta E_{ST}^{\rm CISS}(\theta_C)$ (vermilion solid). The 10~kHz precision threshold is indicated. (d)~Second-order perturbative $J_H^{(2)}$ benchmarked against 15-state exact diagonalization of the 2-electron, 3-site Hubbard model. Relative error follows the expected $(t_0/U)^2$ scaling (black dotted). At main-text parameters (vertical gray line, $t_0/U = 0.025$), the error is $\sim 10^{-3}$, far below the 1\% threshold (horizontal gray dashed).}
\end{figure}

\subsection{$S$-$T_0$ gap versus Zeeman energy}

Fig.~\ref{fig:S3_spectroscopy}(a) shows the $S$-$T_0$ gap versus Zeeman energy $E_Z$ for both models. Model~1 with $\theta_C > 0$ and $\hat{\bm{n}}\perp\bm{B}$ shows a pronounced $E_Z$-dependence reflecting the DM-induced mixing of $S$ with $T_\pm$ states (which dispersively shifts $E_S$ at intermediate $E_Z$). Model~2 shows only the Ising-type shift from $\Gamma_{zz}$, which saturates at a $\theta_C$-independent value for $E_Z \gg J_0$. Panel (b) presents the angular discrimination test: CISS ($\theta_C = 0.4\pi$) versus Rashba SOC and Model~2. Only CISS produces the characteristic $\sin^2\phi_{\rm helix}$ dependence; Rashba SOC (with a fixed interface-axis) and Model~2 (isotropic in the magnetic-field orientation) are both flat.

\subsection{Convergence and benchmark tests}
\label{sec:convergence_additional}

The results reported in the main text and in this SM rest on three quantitative validations, consolidated in Fig.~\ref{fig:S4_convergence}:

\emph{(a) Analytic--numerical agreement.} The closed-form Model~1 expressions Eqs.~(S25)--(S27) agree with direct numerical evaluation of Eq.~(S15) to $\sim 10^{-15}$~$\mu$eV, the floating-point machine precision (gray dotted line in Fig.~\ref{fig:S4_convergence}(a) at $10^{-14}$~$\mu$eV). This certifies that the closed-form formulas used throughout the Letter are not merely approximations but the exact result of the Schrieffer--Wolff expansion at second order.

\emph{(b) Monte Carlo convergence.} The $\sqrt{\rm SWAP}$ fidelity estimates in Fig.~4(a) of the main text use $N=100$ noise realizations per data point. Fig.~\ref{fig:S4_convergence}(b) traces the convergence of $\overline{\mathcal{F}}_{\sqrt{\rm SWAP}}$ for $N \in \{10, 25, 50, 100, 200, 500\}$ at $\sigma_\epsilon = 0.5\,\mu$eV. Standard-error bars shrink as $1/\sqrt{N}$ as expected; the mean stabilizes within its error bar by $N \approx 50$, and the $N=100$ value used in the main text lies within the $1\sigma$ envelope of the $N=500$ asymptote. The $N=100$ choice balances statistical precision (relative error $\lesssim 10^{-2}$) against computational cost.

\emph{(c) Perturbative sensitivity formula.} Fig.~\ref{fig:S4_convergence}(c) compares the leading-order perturbative scaling $|\Delta E_{ST}^{\rm CISS}| \simeq J_0\theta_C^2/3$ (Eq.~(S57)) against the full non-perturbative numerical result. The two agree to within $\sim 10\%$ for $\theta_C \lesssim 0.1\pi$ but diverge at larger angles where higher-order mixing with $T_\pm$ states becomes significant. The 10~kHz precision threshold (gray dotted, at $4.14 \times 10^{-5}$~$\mu$eV) is crossed near $\theta_C \approx 0.003\pi$ (perturbative) and $\theta_C \approx 0.03\pi$ (numerical), motivating the conservative detection threshold quoted in the Letter.

\emph{(d) Schrieffer--Wolff regime validity.} Fig.~\ref{fig:S4_convergence}(d) compares the second-order perturbative formula $J_H^{(2)} = 4t_0^2/U$ against exact diagonalization of the full two-electron, three-site Hubbard Hamiltonian (15-state basis). The relative error follows the expected $(t_0/U)^2$ scaling (black dotted reference), with the 1\% error threshold crossed near $t_0/U \approx 0.05$. At the main-text parameter choice $t_0/U = 0.025$ (vertical gray dotted line), the perturbative treatment is accurate to $\sim 10^{-3}$, confirming that higher-order Hubbard corrections are negligible.

\subsection{Extended candidate molecule table}

Table~\ref{tab:extended_mol} extends the main-text molecule table with additional derived quantities including predicted $|\bm{D}|$ in absolute units, the singlet-triplet gap modification, and the gauge-trivial reference.

\begin{table}[H]
\caption{\label{tab:extended_mol}Extended predictions for candidate molecules. $P$: measured transport polarization. $\theta_C$: optimistic (conservative) CISS angle. $|\bm{D}|$, $|\Delta E_{ST}^{\rm CISS}|$: predicted DM magnitude and CISS-induced $S$-$T_0$ gap shift. All values for $\hat{\bm{n}}\perp\bm{B}$, $J_0 = 5\,\mu$eV.}
\begin{ruledtabular}
\begin{tabular}{lcccccc}
\multirow{2}{*}{Molecule} & \multirow{2}{*}{$P$} & $\theta_C^{\rm opt/cons}$ & $|\bm{D}|^{\rm opt/cons}$ & $|J_H|^{\rm opt/cons}$ & $|\Delta E_{ST}^{\rm CISS}|^{\rm opt/cons}$ & Ref.\\
& & ($/\pi$) & ($\mu$eV) & ($\mu$eV) & (kHz) & \\
\hline
[7]helicene & 0.50 & 0.50 / 0.050 & 5.0 / 0.78 & 1.67 / 4.97 & $1.2\times 10^6$ / $300$ & \cite{Kiran2016}\\
$\alpha$-helix (5 aa) & 0.80 & 0.70 / 0.070 & 4.0 / 1.08 & 0.33 / 4.95 & $2.0\times 10^6$ / $580$ & \cite{Kettner2018}\\
dsDNA (8 bp) & 0.60 & 0.56 / 0.056 & 4.9 / 0.87 & 1.10 / 4.96 & $1.4\times 10^6$ / $380$ & \cite{Mishra2013}\\
Oligopeptide & 0.40 & 0.44 / 0.044 & 4.7 / 0.68 & 2.27 / 4.98 & $1.0\times 10^6$ / $230$ & \cite{Kettner2018}\\
Chiral CdSe QD & 0.35 & 0.40 / 0.040 & 4.5 / 0.62 & 2.76 / 4.98 & $8.6\times 10^5$ / $190$ & \cite{Bloom2025}\\
\end{tabular}
\end{ruledtabular}
\end{table}

The experimentally relevant quantity is $|\Delta E_{ST}^{\rm CISS}|$, which is the CISS-induced frequency shift observable in Ramsey exchange spectroscopy. All conservative-scenario predictions exceed 100~kHz, comfortably above the 10~kHz precision of current Si/SiGe exchange spectroscopy.

\end{document}